\documentclass[%
  preprint,
   amsmath,amssymb,
   aps,
   prb,
  ]{revtex4-1}

\usepackage{graphicx}
\usepackage{color}
\usepackage{bm}% bold math
\usepackage{acro}
\usepackage{xspace}
\usepackage[caption=false]{subfig}
\usepackage[percent]{overpic}

\DeclareAcronym{SM}{
  short=SM,
  long=Supplementary Material,
  }
\DeclareAcronym{BZ}{
  short=BZ,
  long=Brillouin zone,
  }
\DeclareAcronym{TWS}{
  short=TWS,
  long=topological Weyl semimetal
  }
\DeclareAcronym{WS}{
  short=WS,
  long=Weyl semimetal
  }
\DeclareAcronym{TDS}{
  short=TDS,
  long=topological Dirac semimetal
  }
\DeclareAcronym{WN}{
  short=WN,
  long=Weyl node
  }
\DeclareAcronym{3D}{
  short=3D,
  long=three-dimensional,
  }
\DeclareAcronym{2D}{
  short=2D,
  long=two-dimensional,
  }
\DeclareAcronym{SOI}{
  short=SOI,
  long=spin-orbit interaction,
  }
\DeclareAcronym{SOC}{
  short=SOC,
  long=spin-orbit coupling,
  }
\DeclareAcronym{SO}{
  short=SO,
  long=spin-orbit,
  }
\DeclareAcronym{ARPES}{
  short=ARPES,
  long=angular resolved photoemission spectroscopy,
  }
\DeclareAcronym{DFT}{
  short=DFT,
  long=density functional theory,
  }
\DeclareAcronym{XC}{
  short=XC,
  long=exchange and correlation,
  }
\DeclareAcronym{PBE}{
  short=PBE,
  long={Perdew, Burke and Ernzerhof},
  }
\DeclareAcronym{GGA}{
  short=GGA,
  long=generalized-gradient approximation,
  }
\DeclareAcronym{PW}{
  short=PW,
  long=plane wave,
  }
\DeclareAcronym{EELS}{
  short=EELS,
  long=electron energy loss spectra,
  }
\DeclareAcronym{bct}{
  short=bct,
  long=body-centered tetragonal,
  }
\DeclareAcronym{QE}{
  short=QE,
  long=\texttt{Quantum ESPRESSO},
  }
\DeclareAcronym{DOS}{
  short=DOS,
  long=density of states,
  }
\DeclareAcronym{IR}{
  short=IR,
  long=infrared,
  }
\DeclareAcronym{UV}{
  short=UV,
  long=ultraviolet,
  }
\DeclareAcronym{TaAs}{
  short=TaAs,
  long=tantalum arsenide,
  }
\DeclareAcronym{scf}{
  short=scf,
  long=self-consistent,
  }
\DeclareAcronym{HWCC}{
  short=HWCC,
  long=hybrid Wannier charge centers,
  }
\DeclareAcronym{AQHE}{
  short=AQHE,
  long=anomalous quantum Hall effect,
  }

\newcommand{\angstrom}{\textup{\AA}}
\newcommand{\bfk}{{\mathbf k}\xspace}

\begin{document}

  \title{High-throughput screening of Weyl semimetals}

  \author{Davide Grassano}
  \email{davide.grassano@epfl.ch}
  \affiliation{Theory and Simulations of Materials (THEOS), and National Center for Computational Design and Discovery of Novel Materials (MARVEL), \'Ecole Polytechnique F\'ed\'erale de Lausanne, CH-1015 Lausanne, Switzerland}
  
  \author{Nicola Marzari}
  \affiliation{Theory and Simulations of Materials (THEOS), and National Center for Computational Design and Discovery of Novel Materials (MARVEL), \'Ecole Polytechnique F\'ed\'erale de Lausanne, CH-1015 Lausanne, Switzerland}
  \affiliation{Laboratory for Materials Simulations (LMS), Paul Scherrer Institut (PSI), CH-5232, Villigen PSI, Switzerland}

  \author{Davide Campi}
  \affiliation{Theory and Simulations of Materials (THEOS), and National Center for Computational Design and Discovery of Novel Materials (MARVEL), \'Ecole Polytechnique F\'ed\'erale de Lausanne, CH-1015 Lausanne, Switzerland}
  \affiliation{Dipartimento di Scienza dei Materiali, Universita di Milano-Bicocca, Via Cozzi 53, 20125 Milano, Italy}

  %\keywords{Keyword1, Keyword2, Keyword3}

  \begin{abstract}
    Topological Weyl semimetals represent a novel class of non-trivial materials, where band crossings with linear dispersions take place at generic momenta across reciprocal space.
    These crossings give rise to low-energy properties akin to those of Weyl fermions, and are responsible for several exotic phenomena.
    Up to this day, only a handful of Weyl semimetals have been discovered, and the search for new ones remains a very active area.
    The main challenge on the computational side arises from the fact that many of the tools used to identify the topological class of a material do not provide a complete picture in the case of Weyl semimetals.
    In this work, we propose an alternative and inexpensive, criterion to screen for possible Weyl fermions, based on the analysis of the band structure along high-symmetry directions in the absence of spin-orbit coupling.
    We test the method by running a high-throughput screening on a set of 5455 inorganic bulk materials and identify 49 possible candidates for topological properties.
    A further analysis, carried out by identifying and characterizing the crossings in the Brillouin zone, shows us that 3 of these candidates are Weyl semimetals.
    Interestingly, while these 3 materials underwent other high-throughput screenings, none had revealed their topological behavior before.
  \end{abstract}

  \maketitle

  %%%%%%%%%%%%%%%%%%%%%%%%%%%%%%%%%%%%%%%%%%%%%%%%%%%%%%%%%%%%%%%%%%%%%%%%%%%%%%%%%%%%%%%%%%%%%%%%%%
  \section{Introduction}
    Topological materials have emerged in recent years as a very lively area of research in condensed matter physics \cite{wang2017topological}.
    These materials possess unique electronic properties that are not found in conventional materials, such as the presence of gapless helical edge states\cite{dai2008helical} and exotic quasiparticles\cite{barkeshli2013classification,wang2013three}.
    Among them, topological semimetals are of particular interest due to their low energy properties behaving akin to Dirac and Weyl fermions, classifying them as Dirac and \acp{WS}, respectively\cite{armitage2018weyl}.
    The latter are of particular interest as they can manifest several exotic physical properties, such as the Adler-Bell-Jackiw anomaly\cite{adler1969axial,bell1969pcac} which is tied to the observation of negative magneto-transport\cite{ghimire2015magnetotransport,zhang2016signatures,arnold2016negative,gooth2017experimental} and the presence of atypical surface states known as Fermi arcs\cite{Souma.Wang.ea:2016:PRB,Huang.Xu.ea:2015:NC,Xu.Belopolski.ea:2015:S,Xu.Belopolski.ea:2015:SA,Belopolski.Xu.ea:2016:PRL,Xu.Belopolski.ea:2016:PRL}.
    Given this wide range of properties, \acp{WS} have been proposed and used for many applications such as quantum computing\cite{castelvecchi2017strange}, metamaterials\cite{hills2017current} and non-linear optics\cite{oktay2020lasing}.

    Only a handful of \acp{WS} have been currently identified, and the search for new ones is still an open field.
    Computational approaches are very precious in screening for possible candidates \cite{curtarolo2013high}.
    Ideally, in order to classify a material as a \ac{WS}, one would have to look for band crossings across the whole \ac{BZ} and subsequently classify these through methods such as the Wilson loop\cite{yu2011equivalent,alexandradinata2014wilson}.
    Unfortunately, this method is quite expensive as the crossings can be located at any given quasi-momentum in the \ac{BZ}, and hence not suited for high-throughput applications.
    Having complementary methods to predict the topology of a system in a non-resource intensive way would be ideal, before running more expensive checks.
    To this end, many possible methods, not only limited to the discovery of \acp{WS}, are being developed.

    The work of Fu and Kane\cite{fu2007topological} shows that the Z2 topological invariant in inversion symmetric systems can be related to the eigenvalues of symmetry operators at special points in the \ac{BZ} known as "time-reversal invariant momenta" (TRIM) points.
    Following this idea, the connection between the symmetry of a system and its topological invariant has been further explored, leading to the definition of the "symmetry-based indicators"\cite{po2017symmetry,kruthoff2017topological,song2018quantitative,song2018diagnosis,khalaf2018symmetry}, that can be used to predict the topology of a system.
    This approach is relatively inexpensive as it only needs the knowledge of the wave function at high-symmetry points.
    Many screening efforts based on this criteria have been carried out, leading to many novel possible topological materials\cite{zhang2019catalogue,he2019symtopo,vergniory2019complete,tang2019comprehensive}.
    Another proposed way to screen for topological materials is to use the overlap of a wavefunction calculated with and without \ac{SOC} at any point in the \ac{BZ}.
    This leads to the definition of a parameter called spin-orbit spillage\cite{choudhary2019high} which, if larger than a user defined threshold, can indicate that the material is topological.

    While the aforementioned methods have proven to be extremely useful, they still have some shortcomings.
    Symmetry based indicators fail to identify topological semimetals that are not protected by crystal symmetries but only rely on the translational one\cite{gao2020high}.
    For example, \ac{TaAs}\cite{huang2015weyl,lv2015experimental}, a well known \ac{WS}, is not identified by these screenings, as the band inversion that lead to the presence of its \acp{WN} does not happen at a high-symmetry point\cite{zhang2019catalogue}.
    Last, methods based on the spin-orbit spillage can only be applied to materials containing heavy atoms where \ac{SOC} is non-negligible.

    In this work we follow a different route based on the identification of possible fermions in the absence of \ac{SOC}.
    It is known that transition  metal monopnictides are nodal line semimetal when analyzed in absence of \ac{SOC}, while they are \ac{WS} when \ac{SOC} is considered\cite{huang2015weyl,weng2015weyl}.
    In this family of materials, nodal line are present on the mirror planes $k_x = 0$ and $k_y = 0$, leading to the presence of a band crossing near $\Sigma$, along the $\Gamma-\Sigma$ direction in the band structure.
    Each lines decays into 3 pairs of \acp{WN} that drift away slightly from the mirror plane when \ac{SOC} is introduced.
    Starting from this consideration, we decided to screen for materials exhibiting linear or quasi-linear crossings in proximity of the Fermi level in a band structure without \ac{SOC}, as possible candidates for \acp{WS}, to then be confirmed with detailed calculations.

  \section{Methods}
    In this work we rely on \ac{DFT}, carried out using the open-source \ac{PW} \ac{QE} distribution.
    We use two different sets of pseudopotentials: the first ones taken from the {\it SSSP PBE efficiency} (v1.1)\cite{prandini2018precision} library, are used for the initial high-throughput relaxation of primitive cells and atomic positions.
    The second set is taken from the ONCVPSP\cite{hamann2013optimized} library, and is which are used to carry out both non- and fully-relativistic calculations.
    In both case the \ac{XC} functional is given by the \ac{PBE} \ac{GGA}.

    The screening has been carried out starting from structures taken from the ICSD\cite{bergerhoff1987crystallographic,zagorac2019recent} and COD\cite{COD} databases.
    After obtaining the CIF files, the atomic positions and cell parameters have been relaxed with the aforementioned code and protocols, without \ac{SOC}, with a threshold for the forces on the atoms of $4 \cdot 10^{-4}$~Ry/\angstrom.
    For consistency, we then relaxed the position of the atoms again using the ONCVPSP pseudopotentials.
    Successively, the seekpath\cite{hinuma2017band} tool is used in order to generate the high-symmetry paths that are used to compute the band structures (still without \ac{SOC}).
    The resulting band structures are analyzed, screening for linear or quasi-linear crossings within 1~eV from the Fermi level.

    Next, we need to identify if any band crossing within the \ac{BZ} is still present when the \ac{SOC} is introduced, focusing on a neighborhood of the original crossing in the non-\ac{SOC} band structure.
    In order to do so, first we run a \ac{scf} calculation using the same atomic structure, but with increased cutoff and accuracy thresholds, in order to guarantee a numerical precision on the energy eigenvalues of the order of $\sim 0.1$~meV.
    We than proceed to iteratively run a series of non-\ac{scf} calculations with increasingly denser meshes of k-points. 
    The meshes are constructed starting from the "low-gap" points identified in the previous step;
    these are the points in the \ac{BZ} with a gap low enough that, within the the current step size of the mesh, could witness a linear crossing, without assuming a Fermi velocity higher than 3 times that of graphene.
    The procedure is repeated for every point, until they are either discarded by the previous criteria or, if the gap is lower than 2.5~meV, are considered as candidates for the calculation of the Chern number. 
    During this phase, the number of k-points that needs to be computed can vary from a few hundred to several thousand, depending on the amount of trivial points present, which are defined as low-gap points in the band structure not related to the topology of the system.

    The final step is thus to analyze the Chern number of the promising crossings, which should be $\pm 1$ in a \ac{WS}.
    For this purpose we use \texttt{Z2pack}\cite{gresch2017z2pack}, a python tool that works in combination with \ac{QE} (or other \ac{PW} codes) and \texttt{Wannier90}\cite{pizzi2020wannier90}, which implements the calculation of topological invariants based on the evolution of \acp{HWCC} \cite{soluyanov2011computing}.
    We repeat this calculation by increasing the thresholds until all the convergence criteria of \texttt{Z2pack} are satisfied, or until the distance between neighboring lines of k-points needed to calculate the \acp{HWCC} is lower than $10^{-4}$ \angstrom$^{-1}$.

    If all convergence criteria are satisfied and the result for the Chern number is non-zero, we refine the position of the crossing, down to a gap of 0.1~meV.
    We then replicate it, starting from the irreducible \ac{BZ} and applying the crystal symmetries, and calculate the Chern number for all the resulting crossings in order to distinguish the pairs of $\pm$ chirality.
    We also compute the band dispersion for every type of node along the $k_{x/y/z}$ cartesian directions and also along $q_{x/y/z}$, where $q_x$ is the direction connecting two adjacent nodes in a pair, and $q_{y/z}$ are two orthogonal directions.
    For all the Weyl semimetals identified, we also compute the \ac{DOS} and perform a series of \ac{DFT} calculations testing for several possible ferromagnetic and anti-ferromagnetic configurations, with different sizes of supercells and initial spin configurations.

    The entire process has been automated by writing workflows using AiiDA\cite{pizzi2016aiida,huber2020aiida}, which let us keep track of the metadata and provenance for every step of the calculation.
    The relative plugins for \ac{QE} and \texttt{Z2pack} have also been employed.
    All the data produced in this work, including input and output files for the calculations is available on the Materials Cloud \cite{talirz2020materials} at Ref.~\onlinecite{mcloud2023weyl}.

  \section{Results and discussions}
    \begin{figure}
    % \centering
      \begin{tabular}[b]{c}
        \begin{overpic}[width=0.37\textwidth]{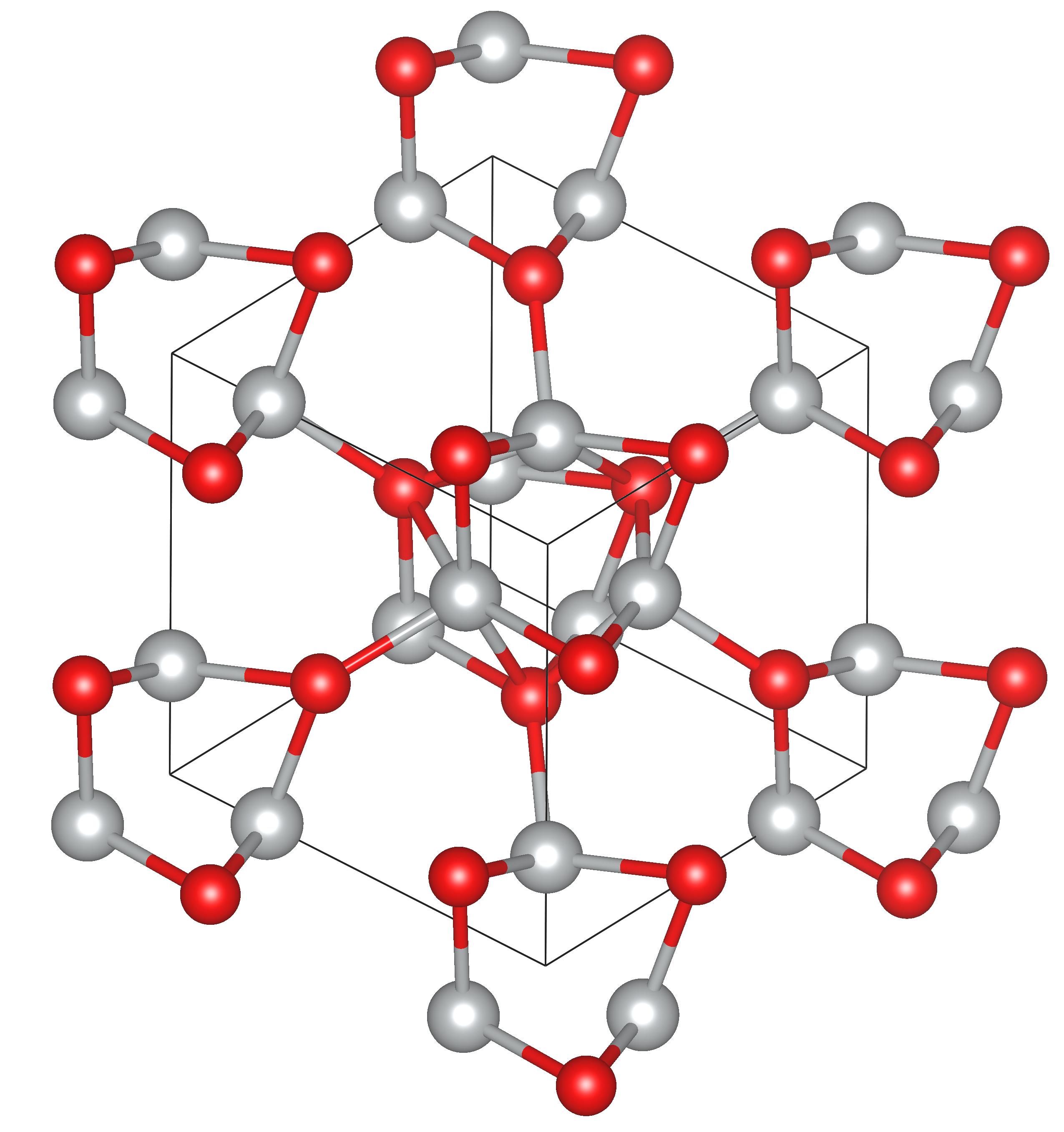}
          \put(0,100){{\bf \LARGE (a)}}
        \end{overpic}
        \\
        \begin{overpic}[width=0.37\textwidth]{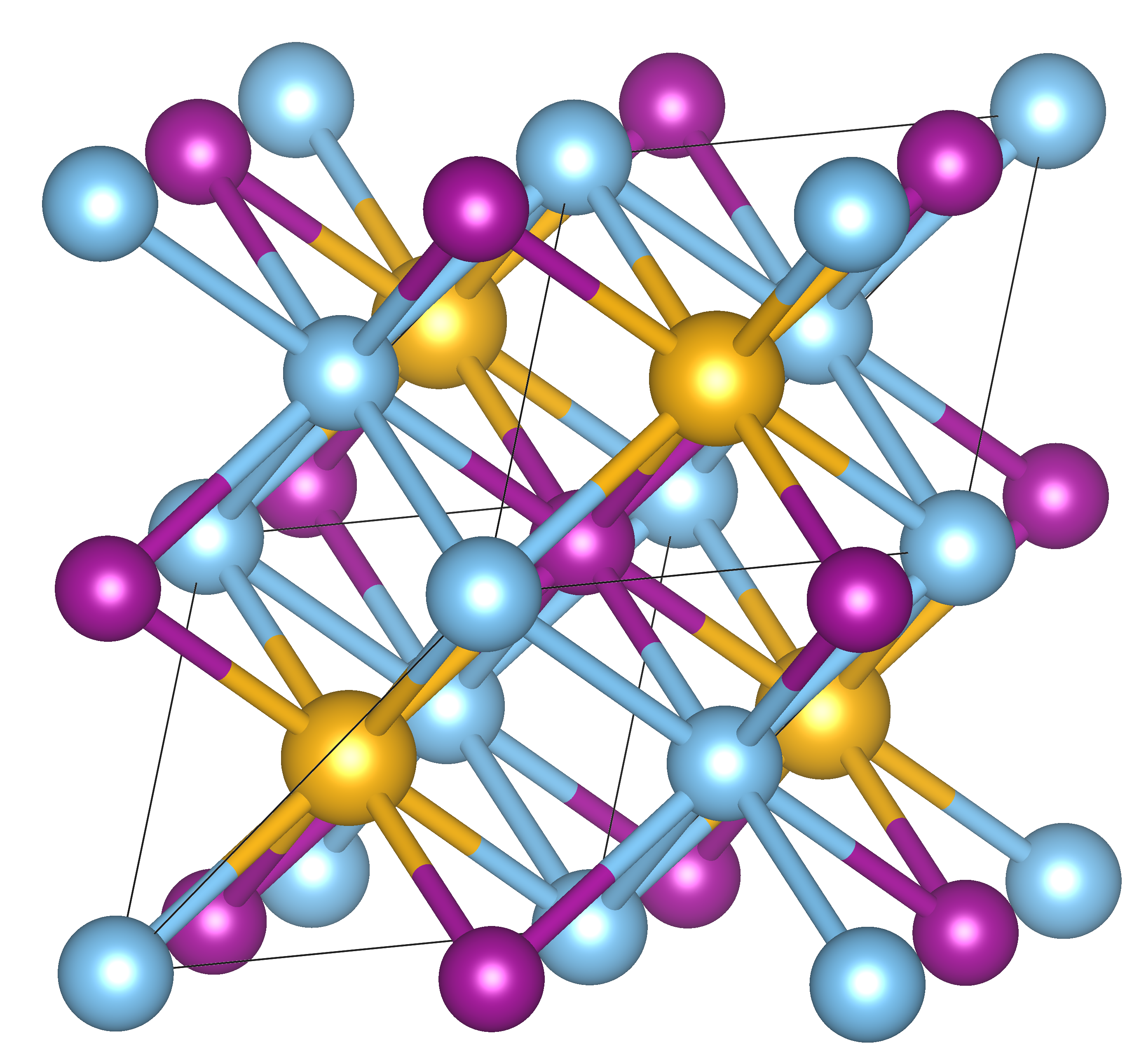}
          \put(0,95){{\bf \LARGE (b)}}
        \end{overpic}
      \end{tabular}
      \caption{
        \label{fig:structures}
        (a) The millerite structure of the two nickel chalcogenides.
        Ni atoms are represented in grey, while S/Se atoms are red.
        (b) The Heusler compound structure of InMnTi$_2$.
        The In, Mn and Ti atoms are shown in purple, orange and cyan respectively.
      }
    \end{figure}
    The screening described in the methods started from 5455 structures including a mix of \ac{2D} layered materials and \ac{3D} materials.
    Out of these, 49 candidates have been identified, having band crossing along high-symmetry directions in the non-\ac{SOC} calculations (see \ac{SM} and Ref.~\onlinecite{mcloud2023weyl}).
    When \ac{SOC} is introduced, the screening for low-gap points reveals that 6 out of the 49 materials exhibit crossings in the \ac{BZ} zone (see \ac{SM}).
    For the remaining ones, a gap opens ranging from a few meV up to 300~meV.
    While these materials are being excluded as candidates for topological semimetals, they could still be non trivial insulators and could warrant additional study.

\begin{figure*}
      \includegraphics[width=0.9\textwidth, trim=0 70 0 50]{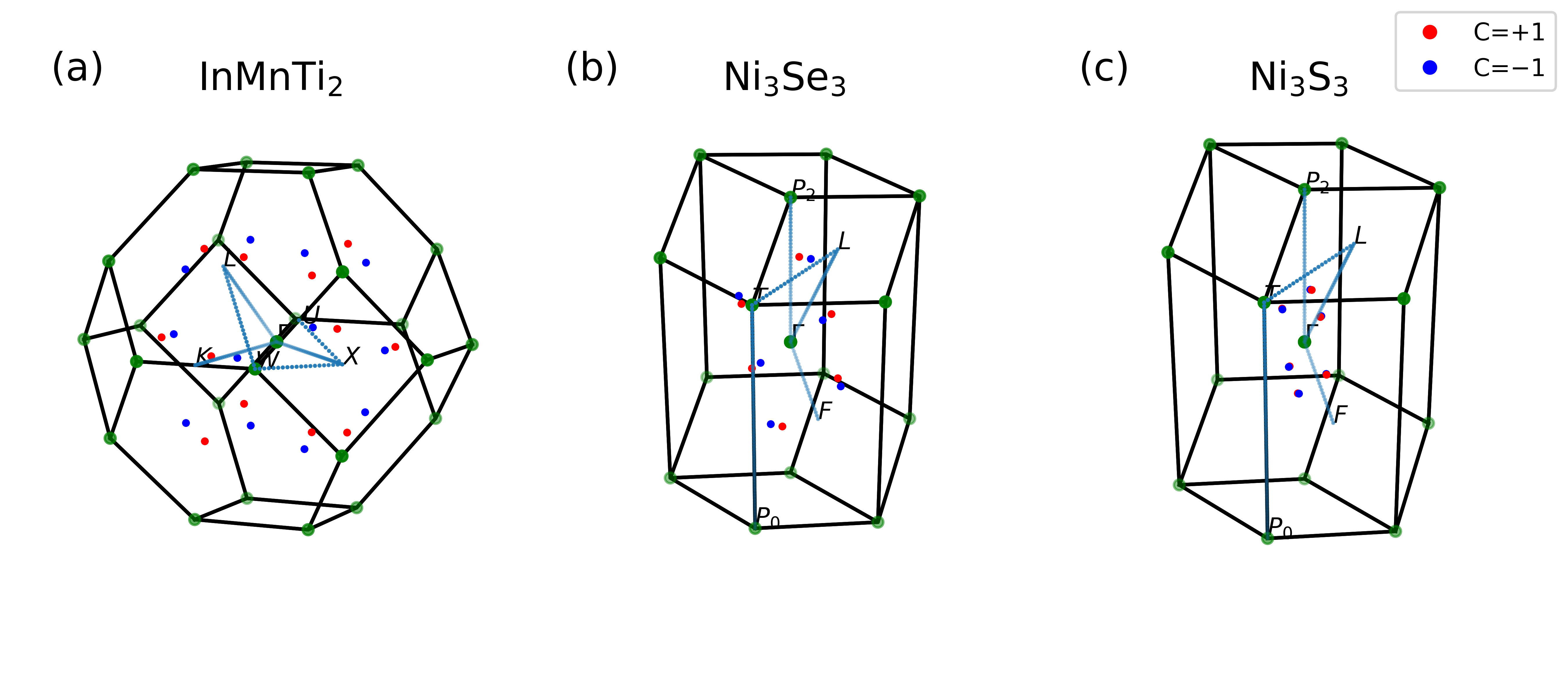}
      \caption{
        \label{fig:crossings}
        Plot of the \acl{BZ} of InMnTi$_2$ (a), Ni$_3$Se$_3$ (b) and  Ni$_3$S$_3$ (c).
        The cyan dots represents the high-symmetry path generate by the seekpath tool.
        The green dots represents the vertex of the \ac{BZ}.
        The red (blue) dots represents the position of the Weyl nodes with chirality +1 (-1).
        The nodes in the same pair of  Ni$_3$S$_3$, that at the scale of this figure overlap with each other.
        K-point with low gap, but zero Chern number are not included for clarity, but are show in the \ac{SM}.
      }
    \end{figure*}

    The computation of the Chern number for the remaining 6 candidates reveals that 3 of them are \acp{WS}.
    The other 3 materials (see \ac{SM}) could be interesting for further studies as possible Dirac semimetals.
    Among the Weyl semimetals identified, two of them, namely Ni$_3$S$_3$ and Ni$_3$Se$_3$, are nickel chalcogenides with a millerite structure\cite{alsen1925rontgenographische} (see Fig.~\ref{fig:structures}a) and space group 160.
    The last one is InMnTi$_2$, a Heusler compound\cite{graf2009crystal} with a Hg$_2$CuTi-type structure (see Fig.~\ref{fig:structures}b) and space group 216.
    Interestingly, a material with an analog structure where In is replaced by aluminum has also been recently identified as a \ac{WS}\cite{shi2018prediction}.
    Given that all the materials identified lack inversion symmetry, we expect to find for each of them a number of \acp{WN} that is a multiple of 4, since for every node in a pair $(\bfk_0, \ \bfk^\prime_0)$, there must exist another one at $(-\bfk_0, \ -\bfk^\prime_0)$.
    Indeed, for 2 candidates we find 12 \acp{WN} across the entire \ac{BZ}, while 24 are found for the remaining one.
    Due to the nature of the screening, we also expect to identify pairs of \acp{WN} close to each other and with opposite chiralities, in proximity of the original non-\ac{SOC} crossings.
    For the two nickel chalcogenides we find a single pair of \acp{WN} in proximity of the $\Gamma - L$ direction, which unfolds into 6 pairs.
    This is in agreement with the findings of Zhang et al\cite{zhang2019catalogue}, where the two nickel chalcogenides have been classified as potential semimetals with crossings on the $\Gamma - L$ line.
    For InMnTi$_2$ one pair of nodes is found close to the $\Gamma - K$ direction, that unfolds into the 24 \acp{WN}.
    The location of the crossing in the \ac{BZ} can be visually observed in Fig.~\ref{fig:crossings}, with the positions given in Table~\ref{tab:nodes_data}.
    For all 3 materials we also find that using standard \ac{DFT} (as opposed to more advanced methods such as DFT+U), none of them exhibit a magnetic ground states.
    \begin{figure*}
      \includegraphics[width=0.9\textwidth]{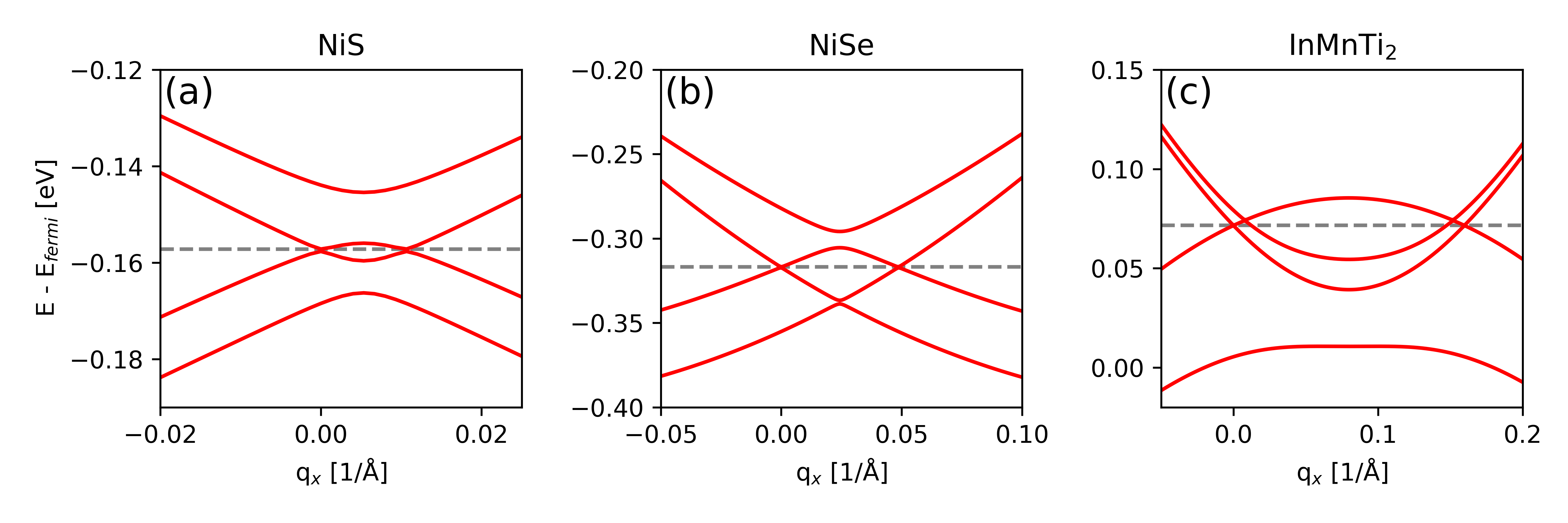}
      \caption{
        \label{fig:qx_dispersion}
        Band dispersions along the line connecting two Weyl nodes in an adjacent pair for (a) Ni$_3$S$_3$, (b) Ni$_3$Se$_3$ and (c) InMnTi$_2$.
        The dashed grey line shows the position of the node in energy with respect to the Fermi level.
        In all panels the coordinate $q_x = 0$ is centered on the left node in a pair an the $q_x$ of the second node denotes the separation in momentum space.
      }
    \end{figure*}
    In Table~\ref{tab:nodes_data} we provide the characterizing feature of the \acp{WN}, with the position of the node in both momentum space (k$_{x/y/z}$) and energy space with respect to the Fermi level (E$_W$) given.
    The closer the latter is to zero, the easier it should be to observe the Weyl-like behavior in the material.
    In addition, by using perturbations such as strain\cite{liu2014tuning}, electric field\cite{Drummond.2012,grassano2018detection}, substrate interactions\cite{menshchikova2013band} or doping\cite{edmonds2016molecular}, it has been shown to be possible to engineer the band structure of a topological material to make its properties more accessible while preserving the topology.
    We also provide the linearity range, in both momentum ($\Delta$K$_{lin}$) and energy space($\Delta$E$_{lin}$), here defined as the range in which a linear fit of both the upper or lower Weyl band, in all directions, can be performed with the square of the correlation coefficient $r$ being greater than $0.98$ (see \ac{SM}).
    This parameter can be used to determine the range of energies in which the low-energy properties are expected to behave accordingly to the Weyl Hamiltonian.
    Finally, we provide the separation between the nodes in momentum space, which is tied to the linearity range of the materials due to the pairing effect\cite{grassano2019influence}, and also relates to the possible strength of the \ac{AQHE}\cite{wang_3d_2017, yang_quantum_2011}.

    Among the materials identified, Ni$_3$S$_3$ cannot be considered as a good candidate for the realization of Weyl physics for several reasons.
    First, the separation of the nodes within a pair is only 0.01~\angstrom$^{-1}$ (see Fig.~\ref{fig:qx_dispersion}a), leading to a linearity range of only 3~meV.
    Second, the node position is also quite far from the Fermi level (-157~meV).
    While this is still within an accessible range, it might be difficult to engineer the band structure without altering the topology of the material, given to how close the nodes are.
    Finally, the nodes are mixed with metallic bands, as can be seen in the \ac{DOS} shown in the \ac{SM}.
    Ni$_3$Se$_3$ behaves similarly to its sulfide counterpart, with some differences due to the greater effect from the \ac{SOC}, leading to an higher node separation of 0.05~\angstrom$^{-1}$ (see Fig.~\ref{fig:qx_dispersion}b).
    The linearity range is also higher and can be observed up to 30~meV away from the node.
    Also for this material, the \acp{WN} are mixed with metallic bands and the nodes are even further away from the Fermi level (-317~meV).
    InMnTi$_2$ seems to be instead an exceedingly promising candidate for the realization of Weyl physics.
    For the only type of pair of \acp{WN} identified, we find a separation of 0.16~\angstrom$^{-1}$, which is more than twice that of the W2 node in TaAs\cite{grassano2018validity}, the one with the highest separation among the transition-metal monopnictides, which also translates into a linearity range of 48 meV.
    The nodes are also close to the Fermi level, being shifted from it only by 72~meV, which is comparable to the case of the transition-metal monopnictides\cite{lee2015fermi}.
    Furthermore, as highlighted by the fact that the \ac{DOS} (see \ac{SM}), goes to zero at E$_W$ and follows a parabolic behavior, the nodes are not mixed with any other band or trivial points, making them easily accessible.
    \begin{table*}
      \caption{
        \label{tab:nodes_data}
        Data obtained from the analysis of the band dispersions at the nodes along the q$_{x/y/z}$ directions.
        First, the position on the nodes in reciprocal space is given;
        E$_W$ represents the distance of the node from the Fermi level, and
        $\Delta _{W \rightarrow W}$ represents the separation between the nodes of a pair.
        $\Delta$K$_{lin}$ and $\Delta$E$_{lin}$ represent the range in which the linearity of the bands can be observed both in momentum and energy space respectively.
      }
      \centering
      \begin{tabular}[b]{|c|ccc|c|c|c|c|}
        \hline
                     & k$_x$ (\angstrom$^{-1}$) 
                     & k$_y$ (\angstrom$^{-1}$) 
                     & k$_z$ (\angstrom$^{-1}$) 
                     & E$_W$ (meV) 
                     & $\Delta _{W \rightarrow W}$ ($10^{-2}$\angstrom$^{-1}$) 
                     & $\Delta$K$_{lin}$ ($10^{-2}$\angstrom$^{-1}$) 
                     & $\Delta$E$_{lin}$ (meV) \\
        \hline
                NiS  &  0.1116 & 0.0687 & 0.2926 & -157 &   1.1 & 0.4 &  3 \\
               NiSe  &  0.2712 & 0.1165 & 0.3967 & -317 &   4.9 & 1.5 & 30 \\
       InMnTi$_{2}$  &  0.4393 & 0.5522 & 0.0000 &   72 &  16.0 & 5.1 & 45 \\
        \hline
      \end{tabular}
    \end{table*}
    Given that the material with structure analogous to InMnTi$_2$ identified by Shi et al.\cite{shi2018prediction} has been determined to have an antiferromagnetic ground state, with magnetic moments on the Mn and Ti atoms, a further study of the ground state with higher level theories, such as \ac{DFT}+U is required InMnTi2\cite{grassano2022prediction}.
    
  \section{Conclusions}
    In this work we have introduced an alternative/complementary method to the currently existing ones, to facilitate the screening for Weyl semimetals.
    As this approach only relies on band structures computed without \acl{SOC}, it is extremely lightweight in nature and can be implemented in high-throughput calculations.
    Of the 5455 materials used as the starting point for our screening, 49 of them qualified as possible topological materials among which 3 of them have been identified as \aclp{WS} and 3 as potential Dirac semimetals.
    For the two nickel chalcogenides, the Weyl semimetal properties might be difficult to access experimentally, given that for Ni$_3$S$_3$ the nodes in a pair are extremely close to each other, and for Ni$_3$Se$_3$ they are far below the Fermi level.
    Nevertheless both materials could be exploited using e.g. an external magnetic field to try and separate the nodes, or with doping to shift the Fermi level.
    Conversely, the Heusler compound InMnTi$_2$ has shown to be a prime candidate for the realization of Weyl physics, as the nodes in a pair are well separated and close to the Fermi level.
    Further theoretical and experimental studies on this material are warranted, in order to search for the presence of Fermi arc and atypical magneto-transport properties.

  \section{Acknowledgements}
    This research was supported by the NCCR MARVEL, a National Centre of Competence in Research, funded by the Swiss National Science Foundation (grant number 182892).
    We acknowledge access to Piz Daint or Eiger@Alps at the Swiss National Supercomputing Centre, Switzerland under the EPFL's share with the project ID mr0

  \bibliography{bibliography}

\begin{thebibliography}{65}%
\makeatletter
\providecommand \@ifxundefined [1]{%
 \@ifx{#1\undefined}
}%
\providecommand \@ifnum [1]{%
 \ifnum #1\expandafter \@firstoftwo
 \else \expandafter \@secondoftwo
 \fi
}%
\providecommand \@ifx [1]{%
 \ifx #1\expandafter \@firstoftwo
 \else \expandafter \@secondoftwo
 \fi
}%
\providecommand \natexlab [1]{#1}%
\providecommand \enquote  [1]{``#1''}%
\providecommand \bibnamefont  [1]{#1}%
\providecommand \bibfnamefont [1]{#1}%
\providecommand \citenamefont [1]{#1}%
\providecommand \href@noop [0]{\@secondoftwo}%
\providecommand \href [0]{\begingroup \@sanitize@url \@href}%
\providecommand \@href[1]{\@@startlink{#1}\@@href}%
\providecommand \@@href[1]{\endgroup#1\@@endlink}%
\providecommand \@sanitize@url [0]{\catcode `\\12\catcode `\$12\catcode
  `\&12\catcode `\#12\catcode `\^12\catcode `\_12\catcode `\%12\relax}%
\providecommand \@@startlink[1]{}%
\providecommand \@@endlink[0]{}%
\providecommand \url  [0]{\begingroup\@sanitize@url \@url }%
\providecommand \@url [1]{\endgroup\@href {#1}{\urlprefix }}%
\providecommand \urlprefix  [0]{URL }%
\providecommand \Eprint [0]{\href }%
\providecommand \doibase [0]{http://dx.doi.org/}%
\providecommand \selectlanguage [0]{\@gobble}%
\providecommand \bibinfo  [0]{\@secondoftwo}%
\providecommand \bibfield  [0]{\@secondoftwo}%
\providecommand \translation [1]{[#1]}%
\providecommand \BibitemOpen [0]{}%
\providecommand \bibitemStop [0]{}%
\providecommand \bibitemNoStop [0]{.\EOS\space}%
\providecommand \EOS [0]{\spacefactor3000\relax}%
\providecommand \BibitemShut  [1]{\csname bibitem#1\endcsname}%
\let\auto@bib@innerbib\@empty
%</preamble>
\bibitem [{\citenamefont {Wang}\ and\ \citenamefont
  {Zhang}(2017)}]{wang2017topological}%
  \BibitemOpen
  \bibfield  {author} {\bibinfo {author} {\bibfnamefont {J.}~\bibnamefont
  {Wang}}\ and\ \bibinfo {author} {\bibfnamefont {S.-C.}\ \bibnamefont
  {Zhang}},\ }\href@noop {} {\bibfield  {journal} {\bibinfo  {journal} {Nature
  materials}\ }\textbf {\bibinfo {volume} {16}},\ \bibinfo {pages} {1062}
  (\bibinfo {year} {2017})}\BibitemShut {NoStop}%
\bibitem [{\citenamefont {Dai}\ \emph {et~al.}(2008)\citenamefont {Dai},
  \citenamefont {Hughes}, \citenamefont {Qi}, \citenamefont {Fang},\ and\
  \citenamefont {Zhang}}]{dai2008helical}%
  \BibitemOpen
  \bibfield  {author} {\bibinfo {author} {\bibfnamefont {X.}~\bibnamefont
  {Dai}}, \bibinfo {author} {\bibfnamefont {T.~L.}\ \bibnamefont {Hughes}},
  \bibinfo {author} {\bibfnamefont {X.-L.}\ \bibnamefont {Qi}}, \bibinfo
  {author} {\bibfnamefont {Z.}~\bibnamefont {Fang}}, \ and\ \bibinfo {author}
  {\bibfnamefont {S.-C.}\ \bibnamefont {Zhang}},\ }\href@noop {} {\bibfield
  {journal} {\bibinfo  {journal} {Physical Review B}\ }\textbf {\bibinfo
  {volume} {77}},\ \bibinfo {pages} {125319} (\bibinfo {year}
  {2008})}\BibitemShut {NoStop}%
\bibitem [{\citenamefont {Barkeshli}\ \emph {et~al.}(2013)\citenamefont
  {Barkeshli}, \citenamefont {Jian},\ and\ \citenamefont
  {Qi}}]{barkeshli2013classification}%
  \BibitemOpen
  \bibfield  {author} {\bibinfo {author} {\bibfnamefont {M.}~\bibnamefont
  {Barkeshli}}, \bibinfo {author} {\bibfnamefont {C.-M.}\ \bibnamefont {Jian}},
  \ and\ \bibinfo {author} {\bibfnamefont {X.-L.}\ \bibnamefont {Qi}},\
  }\href@noop {} {\bibfield  {journal} {\bibinfo  {journal} {Physical Review
  B}\ }\textbf {\bibinfo {volume} {88}},\ \bibinfo {pages} {241103} (\bibinfo
  {year} {2013})}\BibitemShut {NoStop}%
\bibitem [{\citenamefont {Wang}\ \emph {et~al.}(2013)\citenamefont {Wang},
  \citenamefont {Weng}, \citenamefont {Wu}, \citenamefont {Dai},\ and\
  \citenamefont {Fang}}]{wang2013three}%
  \BibitemOpen
  \bibfield  {author} {\bibinfo {author} {\bibfnamefont {Z.}~\bibnamefont
  {Wang}}, \bibinfo {author} {\bibfnamefont {H.}~\bibnamefont {Weng}}, \bibinfo
  {author} {\bibfnamefont {Q.}~\bibnamefont {Wu}}, \bibinfo {author}
  {\bibfnamefont {X.}~\bibnamefont {Dai}}, \ and\ \bibinfo {author}
  {\bibfnamefont {Z.}~\bibnamefont {Fang}},\ }\href@noop {} {\bibfield
  {journal} {\bibinfo  {journal} {Physical Review B}\ }\textbf {\bibinfo
  {volume} {88}},\ \bibinfo {pages} {125427} (\bibinfo {year}
  {2013})}\BibitemShut {NoStop}%
\bibitem [{\citenamefont {Armitage}\ \emph {et~al.}(2018)\citenamefont
  {Armitage}, \citenamefont {Mele},\ and\ \citenamefont
  {Vishwanath}}]{armitage2018weyl}%
  \BibitemOpen
  \bibfield  {author} {\bibinfo {author} {\bibfnamefont {N.}~\bibnamefont
  {Armitage}}, \bibinfo {author} {\bibfnamefont {E.}~\bibnamefont {Mele}}, \
  and\ \bibinfo {author} {\bibfnamefont {A.}~\bibnamefont {Vishwanath}},\
  }\href {https://doi.org/10.1103/RevModPhys.90.015001} {\bibfield  {journal}
  {\bibinfo  {journal} {Reviews of Modern Physics}\ }\textbf {\bibinfo {volume}
  {90}},\ \bibinfo {pages} {015001} (\bibinfo {year} {2018})}\BibitemShut
  {NoStop}%
\bibitem [{\citenamefont {Adler}(1969)}]{adler1969axial}%
  \BibitemOpen
  \bibfield  {author} {\bibinfo {author} {\bibfnamefont {S.~L.}\ \bibnamefont
  {Adler}},\ }\href {https://doi.org/10.1103/PhysRev.177.2426} {\bibfield
  {journal} {\bibinfo  {journal} {Physical Review}\ }\textbf {\bibinfo {volume}
  {177}},\ \bibinfo {pages} {2426} (\bibinfo {year} {1969})}\BibitemShut
  {NoStop}%
\bibitem [{\citenamefont {Bell}\ and\ \citenamefont
  {Jackiw}(1969)}]{bell1969pcac}%
  \BibitemOpen
  \bibfield  {author} {\bibinfo {author} {\bibfnamefont {J.~S.}\ \bibnamefont
  {Bell}}\ and\ \bibinfo {author} {\bibfnamefont {R.}~\bibnamefont {Jackiw}},\
  }\href {https://doi.org/10.1007/BF02823296} {\bibfield  {journal} {\bibinfo
  {journal} {Il Nuovo Cimento A (1965-1970)}\ }\textbf {\bibinfo {volume}
  {60}},\ \bibinfo {pages} {47} (\bibinfo {year} {1969})}\BibitemShut {NoStop}%
\bibitem [{\citenamefont {Ghimire}\ \emph {et~al.}(2015)\citenamefont
  {Ghimire}, \citenamefont {Luo}, \citenamefont {Neupane}, \citenamefont
  {Williams}, \citenamefont {Bauer},\ and\ \citenamefont
  {Ronning}}]{ghimire2015magnetotransport}%
  \BibitemOpen
  \bibfield  {author} {\bibinfo {author} {\bibfnamefont {N.~J.}\ \bibnamefont
  {Ghimire}}, \bibinfo {author} {\bibfnamefont {Y.}~\bibnamefont {Luo}},
  \bibinfo {author} {\bibfnamefont {M.}~\bibnamefont {Neupane}}, \bibinfo
  {author} {\bibfnamefont {D.}~\bibnamefont {Williams}}, \bibinfo {author}
  {\bibfnamefont {E.}~\bibnamefont {Bauer}}, \ and\ \bibinfo {author}
  {\bibfnamefont {F.}~\bibnamefont {Ronning}},\ }\href
  {https://doi.org/10.1088/0953-8984/27/15/152201} {\bibfield  {journal}
  {\bibinfo  {journal} {Journal of Physics: Condensed Matter}\ }\textbf
  {\bibinfo {volume} {27}},\ \bibinfo {pages} {152201} (\bibinfo {year}
  {2015})}\BibitemShut {NoStop}%
\bibitem [{\citenamefont {Zhang}\ \emph {et~al.}(2016)\citenamefont {Zhang},
  \citenamefont {Xu}, \citenamefont {Belopolski}, \citenamefont {Yuan},
  \citenamefont {Lin}, \citenamefont {Tong}, \citenamefont {Bian},
  \citenamefont {Alidoust}, \citenamefont {Lee}, \citenamefont {Huang} \emph
  {et~al.}}]{zhang2016signatures}%
  \BibitemOpen
  \bibfield  {author} {\bibinfo {author} {\bibfnamefont {C.-L.}\ \bibnamefont
  {Zhang}}, \bibinfo {author} {\bibfnamefont {S.-Y.}\ \bibnamefont {Xu}},
  \bibinfo {author} {\bibfnamefont {I.}~\bibnamefont {Belopolski}}, \bibinfo
  {author} {\bibfnamefont {Z.}~\bibnamefont {Yuan}}, \bibinfo {author}
  {\bibfnamefont {Z.}~\bibnamefont {Lin}}, \bibinfo {author} {\bibfnamefont
  {B.}~\bibnamefont {Tong}}, \bibinfo {author} {\bibfnamefont {G.}~\bibnamefont
  {Bian}}, \bibinfo {author} {\bibfnamefont {N.}~\bibnamefont {Alidoust}},
  \bibinfo {author} {\bibfnamefont {C.-C.}\ \bibnamefont {Lee}}, \bibinfo
  {author} {\bibfnamefont {S.-M.}\ \bibnamefont {Huang}},  \emph {et~al.},\
  }\href {https://doi.org/10.1038/ncomms10735} {\bibfield  {journal} {\bibinfo
  {journal} {Nature communications}\ }\textbf {\bibinfo {volume} {7}},\
  \bibinfo {pages} {10735} (\bibinfo {year} {2016})}\BibitemShut {NoStop}%
\bibitem [{\citenamefont {Arnold}\ \emph {et~al.}(2016)\citenamefont {Arnold},
  \citenamefont {Shekhar}, \citenamefont {Wu}, \citenamefont {Sun},
  \citenamefont {Dos~Reis}, \citenamefont {Kumar}, \citenamefont {Naumann},
  \citenamefont {Ajeesh}, \citenamefont {Schmidt}, \citenamefont {Grushin}
  \emph {et~al.}}]{arnold2016negative}%
  \BibitemOpen
  \bibfield  {author} {\bibinfo {author} {\bibfnamefont {F.}~\bibnamefont
  {Arnold}}, \bibinfo {author} {\bibfnamefont {C.}~\bibnamefont {Shekhar}},
  \bibinfo {author} {\bibfnamefont {S.-C.}\ \bibnamefont {Wu}}, \bibinfo
  {author} {\bibfnamefont {Y.}~\bibnamefont {Sun}}, \bibinfo {author}
  {\bibfnamefont {R.~D.}\ \bibnamefont {Dos~Reis}}, \bibinfo {author}
  {\bibfnamefont {N.}~\bibnamefont {Kumar}}, \bibinfo {author} {\bibfnamefont
  {M.}~\bibnamefont {Naumann}}, \bibinfo {author} {\bibfnamefont {M.~O.}\
  \bibnamefont {Ajeesh}}, \bibinfo {author} {\bibfnamefont {M.}~\bibnamefont
  {Schmidt}}, \bibinfo {author} {\bibfnamefont {A.~G.}\ \bibnamefont
  {Grushin}},  \emph {et~al.},\ }\href {https://doi.org/10.1038/ncomms11615}
  {\bibfield  {journal} {\bibinfo  {journal} {Nature communications}\ }\textbf
  {\bibinfo {volume} {7}},\ \bibinfo {pages} {11615} (\bibinfo {year}
  {2016})}\BibitemShut {NoStop}%
\bibitem [{\citenamefont {Gooth}\ \emph {et~al.}(2017)\citenamefont {Gooth},
  \citenamefont {Niemann}, \citenamefont {Meng}, \citenamefont {Grushin},
  \citenamefont {Landsteiner}, \citenamefont {Gotsmann}, \citenamefont
  {Menges}, \citenamefont {Schmidt}, \citenamefont {Shekhar}, \citenamefont
  {S{\"u}{\ss}} \emph {et~al.}}]{gooth2017experimental}%
  \BibitemOpen
  \bibfield  {author} {\bibinfo {author} {\bibfnamefont {J.}~\bibnamefont
  {Gooth}}, \bibinfo {author} {\bibfnamefont {A.~C.}\ \bibnamefont {Niemann}},
  \bibinfo {author} {\bibfnamefont {T.}~\bibnamefont {Meng}}, \bibinfo {author}
  {\bibfnamefont {A.~G.}\ \bibnamefont {Grushin}}, \bibinfo {author}
  {\bibfnamefont {K.}~\bibnamefont {Landsteiner}}, \bibinfo {author}
  {\bibfnamefont {B.}~\bibnamefont {Gotsmann}}, \bibinfo {author}
  {\bibfnamefont {F.}~\bibnamefont {Menges}}, \bibinfo {author} {\bibfnamefont
  {M.}~\bibnamefont {Schmidt}}, \bibinfo {author} {\bibfnamefont
  {C.}~\bibnamefont {Shekhar}}, \bibinfo {author} {\bibfnamefont
  {V.}~\bibnamefont {S{\"u}{\ss}}},  \emph {et~al.},\ }\href
  {https://doi.org/10.1038/srep43394} {\bibfield  {journal} {\bibinfo
  {journal} {Nature}\ }\textbf {\bibinfo {volume} {547}},\ \bibinfo {pages}
  {324} (\bibinfo {year} {2017})}\BibitemShut {NoStop}%
\bibitem [{\citenamefont {Souma}\ \emph {et~al.}(2016)\citenamefont {Souma},
  \citenamefont {Wang}, \citenamefont {Kotaka}, \citenamefont {Sato},
  \citenamefont {Nakayama}, \citenamefont {Tanaka}, \citenamefont {Kimizuka},
  \citenamefont {Takahashi}, \citenamefont {Yamauchi}, \citenamefont {Oguchi}
  \emph {et~al.}}]{Souma.Wang.ea:2016:PRB}%
  \BibitemOpen
  \bibfield  {author} {\bibinfo {author} {\bibfnamefont {S.}~\bibnamefont
  {Souma}}, \bibinfo {author} {\bibfnamefont {Z.}~\bibnamefont {Wang}},
  \bibinfo {author} {\bibfnamefont {H.}~\bibnamefont {Kotaka}}, \bibinfo
  {author} {\bibfnamefont {T.}~\bibnamefont {Sato}}, \bibinfo {author}
  {\bibfnamefont {K.}~\bibnamefont {Nakayama}}, \bibinfo {author}
  {\bibfnamefont {Y.}~\bibnamefont {Tanaka}}, \bibinfo {author} {\bibfnamefont
  {H.}~\bibnamefont {Kimizuka}}, \bibinfo {author} {\bibfnamefont
  {T.}~\bibnamefont {Takahashi}}, \bibinfo {author} {\bibfnamefont
  {K.}~\bibnamefont {Yamauchi}}, \bibinfo {author} {\bibfnamefont
  {T.}~\bibnamefont {Oguchi}},  \emph {et~al.},\ }\href@noop {} {\bibfield
  {journal} {\bibinfo  {journal} {Physical Review B}\ }\textbf {\bibinfo
  {volume} {93}},\ \bibinfo {pages} {161112} (\bibinfo {year}
  {2016})}\BibitemShut {NoStop}%
\bibitem [{\citenamefont {Huang}\ \emph
  {et~al.}(2015{\natexlab{a}})\citenamefont {Huang}, \citenamefont {Xu},
  \citenamefont {Belopolski}, \citenamefont {Lee}, \citenamefont {Chang},
  \citenamefont {Wang}, \citenamefont {Alidoust}, \citenamefont {Bian},
  \citenamefont {Neupane}, \citenamefont {Zhang} \emph
  {et~al.}}]{Huang.Xu.ea:2015:NC}%
  \BibitemOpen
  \bibfield  {author} {\bibinfo {author} {\bibfnamefont {S.-M.}\ \bibnamefont
  {Huang}}, \bibinfo {author} {\bibfnamefont {S.-Y.}\ \bibnamefont {Xu}},
  \bibinfo {author} {\bibfnamefont {I.}~\bibnamefont {Belopolski}}, \bibinfo
  {author} {\bibfnamefont {C.-C.}\ \bibnamefont {Lee}}, \bibinfo {author}
  {\bibfnamefont {G.}~\bibnamefont {Chang}}, \bibinfo {author} {\bibfnamefont
  {B.}~\bibnamefont {Wang}}, \bibinfo {author} {\bibfnamefont {N.}~\bibnamefont
  {Alidoust}}, \bibinfo {author} {\bibfnamefont {G.}~\bibnamefont {Bian}},
  \bibinfo {author} {\bibfnamefont {M.}~\bibnamefont {Neupane}}, \bibinfo
  {author} {\bibfnamefont {C.}~\bibnamefont {Zhang}},  \emph {et~al.},\
  }\href@noop {} {\bibfield  {journal} {\bibinfo  {journal} {Nature
  communications}\ }\textbf {\bibinfo {volume} {6}},\ \bibinfo {pages} {7373}
  (\bibinfo {year} {2015}{\natexlab{a}})}\BibitemShut {NoStop}%
\bibitem [{\citenamefont {Xu}\ \emph {et~al.}(2015{\natexlab{a}})\citenamefont
  {Xu}, \citenamefont {Belopolski}, \citenamefont {Alidoust}, \citenamefont
  {Neupane}, \citenamefont {Bian}, \citenamefont {Zhang}, \citenamefont
  {Sankar}, \citenamefont {Chang}, \citenamefont {Yuan}, \citenamefont {Lee}
  \emph {et~al.}}]{Xu.Belopolski.ea:2015:S}%
  \BibitemOpen
  \bibfield  {author} {\bibinfo {author} {\bibfnamefont {S.-Y.}\ \bibnamefont
  {Xu}}, \bibinfo {author} {\bibfnamefont {I.}~\bibnamefont {Belopolski}},
  \bibinfo {author} {\bibfnamefont {N.}~\bibnamefont {Alidoust}}, \bibinfo
  {author} {\bibfnamefont {M.}~\bibnamefont {Neupane}}, \bibinfo {author}
  {\bibfnamefont {G.}~\bibnamefont {Bian}}, \bibinfo {author} {\bibfnamefont
  {C.}~\bibnamefont {Zhang}}, \bibinfo {author} {\bibfnamefont
  {R.}~\bibnamefont {Sankar}}, \bibinfo {author} {\bibfnamefont
  {G.}~\bibnamefont {Chang}}, \bibinfo {author} {\bibfnamefont
  {Z.}~\bibnamefont {Yuan}}, \bibinfo {author} {\bibfnamefont {C.-C.}\
  \bibnamefont {Lee}},  \emph {et~al.},\ }\href
  {https://doi.org/10.1126%2Fscience.aaa9297} {\bibfield  {journal} {\bibinfo
  {journal} {Science}\ }\textbf {\bibinfo {volume} {349}},\ \bibinfo {pages}
  {613} (\bibinfo {year} {2015}{\natexlab{a}})}\BibitemShut {NoStop}%
\bibitem [{\citenamefont {Xu}\ \emph {et~al.}(2015{\natexlab{b}})\citenamefont
  {Xu}, \citenamefont {Belopolski}, \citenamefont {Sanchez}, \citenamefont
  {Zhang}, \citenamefont {Chang}, \citenamefont {Guo}, \citenamefont {Bian},
  \citenamefont {Yuan}, \citenamefont {Lu}, \citenamefont {Chang} \emph
  {et~al.}}]{Xu.Belopolski.ea:2015:SA}%
  \BibitemOpen
  \bibfield  {author} {\bibinfo {author} {\bibfnamefont {S.-Y.}\ \bibnamefont
  {Xu}}, \bibinfo {author} {\bibfnamefont {I.}~\bibnamefont {Belopolski}},
  \bibinfo {author} {\bibfnamefont {D.~S.}\ \bibnamefont {Sanchez}}, \bibinfo
  {author} {\bibfnamefont {C.}~\bibnamefont {Zhang}}, \bibinfo {author}
  {\bibfnamefont {G.}~\bibnamefont {Chang}}, \bibinfo {author} {\bibfnamefont
  {C.}~\bibnamefont {Guo}}, \bibinfo {author} {\bibfnamefont {G.}~\bibnamefont
  {Bian}}, \bibinfo {author} {\bibfnamefont {Z.}~\bibnamefont {Yuan}}, \bibinfo
  {author} {\bibfnamefont {H.}~\bibnamefont {Lu}}, \bibinfo {author}
  {\bibfnamefont {T.-R.}\ \bibnamefont {Chang}},  \emph {et~al.},\ }\href
  {https://dx.doi.org/10.1126%2Fsciadv.1501092} {\bibfield  {journal} {\bibinfo
   {journal} {Science Advances}\ }\textbf {\bibinfo {volume} {1}},\ \bibinfo
  {pages} {e1501092} (\bibinfo {year} {2015}{\natexlab{b}})}\BibitemShut
  {NoStop}%
\bibitem [{\citenamefont {Belopolski}\ \emph {et~al.}(2016)\citenamefont
  {Belopolski}, \citenamefont {Xu}, \citenamefont {Sanchez}, \citenamefont
  {Chang}, \citenamefont {Guo}, \citenamefont {Neupane}, \citenamefont {Zheng},
  \citenamefont {Lee}, \citenamefont {Huang}, \citenamefont {Bian} \emph
  {et~al.}}]{Belopolski.Xu.ea:2016:PRL}%
  \BibitemOpen
  \bibfield  {author} {\bibinfo {author} {\bibfnamefont {I.}~\bibnamefont
  {Belopolski}}, \bibinfo {author} {\bibfnamefont {S.-Y.}\ \bibnamefont {Xu}},
  \bibinfo {author} {\bibfnamefont {D.~S.}\ \bibnamefont {Sanchez}}, \bibinfo
  {author} {\bibfnamefont {G.}~\bibnamefont {Chang}}, \bibinfo {author}
  {\bibfnamefont {C.}~\bibnamefont {Guo}}, \bibinfo {author} {\bibfnamefont
  {M.}~\bibnamefont {Neupane}}, \bibinfo {author} {\bibfnamefont
  {H.}~\bibnamefont {Zheng}}, \bibinfo {author} {\bibfnamefont {C.-C.}\
  \bibnamefont {Lee}}, \bibinfo {author} {\bibfnamefont {S.-M.}\ \bibnamefont
  {Huang}}, \bibinfo {author} {\bibfnamefont {G.}~\bibnamefont {Bian}},  \emph
  {et~al.},\ }\href {https://doi.org/10.1103%2Fphysrevlett.116.066802}
  {\bibfield  {journal} {\bibinfo  {journal} {Physical Review Letters}\
  }\textbf {\bibinfo {volume} {116}},\ \bibinfo {pages} {066802} (\bibinfo
  {year} {2016})}\BibitemShut {NoStop}%
\bibitem [{\citenamefont {Xu}\ \emph {et~al.}(2016)\citenamefont {Xu},
  \citenamefont {Belopolski}, \citenamefont {Sanchez}, \citenamefont {Neupane},
  \citenamefont {Chang}, \citenamefont {Yaji}, \citenamefont {Yuan},
  \citenamefont {Zhang}, \citenamefont {Kuroda}, \citenamefont {Bian} \emph
  {et~al.}}]{Xu.Belopolski.ea:2016:PRL}%
  \BibitemOpen
  \bibfield  {author} {\bibinfo {author} {\bibfnamefont {S.-Y.}\ \bibnamefont
  {Xu}}, \bibinfo {author} {\bibfnamefont {I.}~\bibnamefont {Belopolski}},
  \bibinfo {author} {\bibfnamefont {D.~S.}\ \bibnamefont {Sanchez}}, \bibinfo
  {author} {\bibfnamefont {M.}~\bibnamefont {Neupane}}, \bibinfo {author}
  {\bibfnamefont {G.}~\bibnamefont {Chang}}, \bibinfo {author} {\bibfnamefont
  {K.}~\bibnamefont {Yaji}}, \bibinfo {author} {\bibfnamefont {Z.}~\bibnamefont
  {Yuan}}, \bibinfo {author} {\bibfnamefont {C.}~\bibnamefont {Zhang}},
  \bibinfo {author} {\bibfnamefont {K.}~\bibnamefont {Kuroda}}, \bibinfo
  {author} {\bibfnamefont {G.}~\bibnamefont {Bian}},  \emph {et~al.},\ }\href
  {https://doi.org/10.1103%2Fphysrevlett.116.096801} {\bibfield  {journal}
  {\bibinfo  {journal} {Physical Review Letters}\ }\textbf {\bibinfo {volume}
  {116}},\ \bibinfo {pages} {096801} (\bibinfo {year} {2016})}\BibitemShut
  {NoStop}%
\bibitem [{\citenamefont {Castelvecchi}(2017)}]{castelvecchi2017strange}%
  \BibitemOpen
  \bibfield  {author} {\bibinfo {author} {\bibfnamefont {D.}~\bibnamefont
  {Castelvecchi}},\ }\href@noop {} {\bibfield  {journal} {\bibinfo  {journal}
  {Nature News}\ }\textbf {\bibinfo {volume} {547}},\ \bibinfo {pages} {272}
  (\bibinfo {year} {2017})}\BibitemShut {NoStop}%
\bibitem [{\citenamefont {Hills}\ \emph {et~al.}(2017)\citenamefont {Hills},
  \citenamefont {Kusmartseva},\ and\ \citenamefont
  {Kusmartsev}}]{hills2017current}%
  \BibitemOpen
  \bibfield  {author} {\bibinfo {author} {\bibfnamefont {R.~D.}\ \bibnamefont
  {Hills}}, \bibinfo {author} {\bibfnamefont {A.}~\bibnamefont {Kusmartseva}},
  \ and\ \bibinfo {author} {\bibfnamefont {F.}~\bibnamefont {Kusmartsev}},\
  }\href {https://doi.org/10.1103/PhysRevB.95.214103} {\bibfield  {journal}
  {\bibinfo  {journal} {Physical Review B}\ }\textbf {\bibinfo {volume} {95}},\
  \bibinfo {pages} {214103} (\bibinfo {year} {2017})}\BibitemShut {NoStop}%
\bibitem [{\citenamefont {Oktay}\ \emph {et~al.}(2020)\citenamefont {Oktay},
  \citenamefont {Sar{\i}saman},\ and\ \citenamefont {Tas}}]{oktay2020lasing}%
  \BibitemOpen
  \bibfield  {author} {\bibinfo {author} {\bibfnamefont {G.}~\bibnamefont
  {Oktay}}, \bibinfo {author} {\bibfnamefont {M.}~\bibnamefont {Sar{\i}saman}},
  \ and\ \bibinfo {author} {\bibfnamefont {M.}~\bibnamefont {Tas}},\
  }\href@noop {} {\bibfield  {journal} {\bibinfo  {journal} {Scientific
  Reports}\ }\textbf {\bibinfo {volume} {10}},\ \bibinfo {pages} {1} (\bibinfo
  {year} {2020})}\BibitemShut {NoStop}%
\bibitem [{\citenamefont {Curtarolo}\ \emph {et~al.}(2013)\citenamefont
  {Curtarolo}, \citenamefont {Hart}, \citenamefont {Nardelli}, \citenamefont
  {Mingo}, \citenamefont {Sanvito},\ and\ \citenamefont
  {Levy}}]{curtarolo2013high}%
  \BibitemOpen
  \bibfield  {author} {\bibinfo {author} {\bibfnamefont {S.}~\bibnamefont
  {Curtarolo}}, \bibinfo {author} {\bibfnamefont {G.~L.}\ \bibnamefont {Hart}},
  \bibinfo {author} {\bibfnamefont {M.~B.}\ \bibnamefont {Nardelli}}, \bibinfo
  {author} {\bibfnamefont {N.}~\bibnamefont {Mingo}}, \bibinfo {author}
  {\bibfnamefont {S.}~\bibnamefont {Sanvito}}, \ and\ \bibinfo {author}
  {\bibfnamefont {O.}~\bibnamefont {Levy}},\ }\href@noop {} {\bibfield
  {journal} {\bibinfo  {journal} {Nature materials}\ }\textbf {\bibinfo
  {volume} {12}},\ \bibinfo {pages} {191} (\bibinfo {year} {2013})}\BibitemShut
  {NoStop}%
\bibitem [{\citenamefont {Yu}\ \emph {et~al.}(2011)\citenamefont {Yu},
  \citenamefont {Qi}, \citenamefont {Bernevig}, \citenamefont {Fang},\ and\
  \citenamefont {Dai}}]{yu2011equivalent}%
  \BibitemOpen
  \bibfield  {author} {\bibinfo {author} {\bibfnamefont {R.}~\bibnamefont
  {Yu}}, \bibinfo {author} {\bibfnamefont {X.~L.}\ \bibnamefont {Qi}}, \bibinfo
  {author} {\bibfnamefont {A.}~\bibnamefont {Bernevig}}, \bibinfo {author}
  {\bibfnamefont {Z.}~\bibnamefont {Fang}}, \ and\ \bibinfo {author}
  {\bibfnamefont {X.}~\bibnamefont {Dai}},\ }\href@noop {} {\bibfield
  {journal} {\bibinfo  {journal} {Physical Review B}\ }\textbf {\bibinfo
  {volume} {84}},\ \bibinfo {pages} {075119} (\bibinfo {year}
  {2011})}\BibitemShut {NoStop}%
\bibitem [{\citenamefont {Alexandradinata}\ \emph {et~al.}(2014)\citenamefont
  {Alexandradinata}, \citenamefont {Dai},\ and\ \citenamefont
  {Bernevig}}]{alexandradinata2014wilson}%
  \BibitemOpen
  \bibfield  {author} {\bibinfo {author} {\bibfnamefont {A.}~\bibnamefont
  {Alexandradinata}}, \bibinfo {author} {\bibfnamefont {X.}~\bibnamefont
  {Dai}}, \ and\ \bibinfo {author} {\bibfnamefont {B.~A.}\ \bibnamefont
  {Bernevig}},\ }\href@noop {} {\bibfield  {journal} {\bibinfo  {journal}
  {Physical Review B}\ }\textbf {\bibinfo {volume} {89}},\ \bibinfo {pages}
  {155114} (\bibinfo {year} {2014})}\BibitemShut {NoStop}%
\bibitem [{\citenamefont {Fu}\ and\ \citenamefont
  {Kane}(2007)}]{fu2007topological}%
  \BibitemOpen
  \bibfield  {author} {\bibinfo {author} {\bibfnamefont {L.}~\bibnamefont
  {Fu}}\ and\ \bibinfo {author} {\bibfnamefont {C.~L.}\ \bibnamefont {Kane}},\
  }\href@noop {} {\bibfield  {journal} {\bibinfo  {journal} {Physical Review
  B}\ }\textbf {\bibinfo {volume} {76}},\ \bibinfo {pages} {045302} (\bibinfo
  {year} {2007})}\BibitemShut {NoStop}%
\bibitem [{\citenamefont {Po}\ \emph {et~al.}(2017)\citenamefont {Po},
  \citenamefont {Vishwanath},\ and\ \citenamefont {Watanabe}}]{po2017symmetry}%
  \BibitemOpen
  \bibfield  {author} {\bibinfo {author} {\bibfnamefont {H.~C.}\ \bibnamefont
  {Po}}, \bibinfo {author} {\bibfnamefont {A.}~\bibnamefont {Vishwanath}}, \
  and\ \bibinfo {author} {\bibfnamefont {H.}~\bibnamefont {Watanabe}},\
  }\href@noop {} {\bibfield  {journal} {\bibinfo  {journal} {Nature
  communications}\ }\textbf {\bibinfo {volume} {8}},\ \bibinfo {pages} {1}
  (\bibinfo {year} {2017})}\BibitemShut {NoStop}%
\bibitem [{\citenamefont {Kruthoff}\ \emph {et~al.}(2017)\citenamefont
  {Kruthoff}, \citenamefont {De~Boer}, \citenamefont {Van~Wezel}, \citenamefont
  {Kane},\ and\ \citenamefont {Slager}}]{kruthoff2017topological}%
  \BibitemOpen
  \bibfield  {author} {\bibinfo {author} {\bibfnamefont {J.}~\bibnamefont
  {Kruthoff}}, \bibinfo {author} {\bibfnamefont {J.}~\bibnamefont {De~Boer}},
  \bibinfo {author} {\bibfnamefont {J.}~\bibnamefont {Van~Wezel}}, \bibinfo
  {author} {\bibfnamefont {C.~L.}\ \bibnamefont {Kane}}, \ and\ \bibinfo
  {author} {\bibfnamefont {R.-J.}\ \bibnamefont {Slager}},\ }\href@noop {}
  {\bibfield  {journal} {\bibinfo  {journal} {Physical Review X}\ }\textbf
  {\bibinfo {volume} {7}},\ \bibinfo {pages} {041069} (\bibinfo {year}
  {2017})}\BibitemShut {NoStop}%
\bibitem [{\citenamefont {Song}\ \emph
  {et~al.}(2018{\natexlab{a}})\citenamefont {Song}, \citenamefont {Zhang},
  \citenamefont {Fang},\ and\ \citenamefont {Fang}}]{song2018quantitative}%
  \BibitemOpen
  \bibfield  {author} {\bibinfo {author} {\bibfnamefont {Z.}~\bibnamefont
  {Song}}, \bibinfo {author} {\bibfnamefont {T.}~\bibnamefont {Zhang}},
  \bibinfo {author} {\bibfnamefont {Z.}~\bibnamefont {Fang}}, \ and\ \bibinfo
  {author} {\bibfnamefont {C.}~\bibnamefont {Fang}},\ }\href@noop {} {\bibfield
   {journal} {\bibinfo  {journal} {Nature communications}\ }\textbf {\bibinfo
  {volume} {9}},\ \bibinfo {pages} {1} (\bibinfo {year}
  {2018}{\natexlab{a}})}\BibitemShut {NoStop}%
\bibitem [{\citenamefont {Song}\ \emph
  {et~al.}(2018{\natexlab{b}})\citenamefont {Song}, \citenamefont {Zhang},\
  and\ \citenamefont {Fang}}]{song2018diagnosis}%
  \BibitemOpen
  \bibfield  {author} {\bibinfo {author} {\bibfnamefont {Z.}~\bibnamefont
  {Song}}, \bibinfo {author} {\bibfnamefont {T.}~\bibnamefont {Zhang}}, \ and\
  \bibinfo {author} {\bibfnamefont {C.}~\bibnamefont {Fang}},\ }\href@noop {}
  {\bibfield  {journal} {\bibinfo  {journal} {Physical Review X}\ }\textbf
  {\bibinfo {volume} {8}},\ \bibinfo {pages} {031069} (\bibinfo {year}
  {2018}{\natexlab{b}})}\BibitemShut {NoStop}%
\bibitem [{\citenamefont {Khalaf}\ \emph {et~al.}(2018)\citenamefont {Khalaf},
  \citenamefont {Po}, \citenamefont {Vishwanath},\ and\ \citenamefont
  {Watanabe}}]{khalaf2018symmetry}%
  \BibitemOpen
  \bibfield  {author} {\bibinfo {author} {\bibfnamefont {E.}~\bibnamefont
  {Khalaf}}, \bibinfo {author} {\bibfnamefont {H.~C.}\ \bibnamefont {Po}},
  \bibinfo {author} {\bibfnamefont {A.}~\bibnamefont {Vishwanath}}, \ and\
  \bibinfo {author} {\bibfnamefont {H.}~\bibnamefont {Watanabe}},\ }\href@noop
  {} {\bibfield  {journal} {\bibinfo  {journal} {Physical Review X}\ }\textbf
  {\bibinfo {volume} {8}},\ \bibinfo {pages} {031070} (\bibinfo {year}
  {2018})}\BibitemShut {NoStop}%
\bibitem [{\citenamefont {Zhang}\ \emph {et~al.}(2019)\citenamefont {Zhang},
  \citenamefont {Jiang}, \citenamefont {Song}, \citenamefont {Huang},
  \citenamefont {He}, \citenamefont {Fang}, \citenamefont {Weng},\ and\
  \citenamefont {Fang}}]{zhang2019catalogue}%
  \BibitemOpen
  \bibfield  {author} {\bibinfo {author} {\bibfnamefont {T.}~\bibnamefont
  {Zhang}}, \bibinfo {author} {\bibfnamefont {Y.}~\bibnamefont {Jiang}},
  \bibinfo {author} {\bibfnamefont {Z.}~\bibnamefont {Song}}, \bibinfo {author}
  {\bibfnamefont {H.}~\bibnamefont {Huang}}, \bibinfo {author} {\bibfnamefont
  {Y.}~\bibnamefont {He}}, \bibinfo {author} {\bibfnamefont {Z.}~\bibnamefont
  {Fang}}, \bibinfo {author} {\bibfnamefont {H.}~\bibnamefont {Weng}}, \ and\
  \bibinfo {author} {\bibfnamefont {C.}~\bibnamefont {Fang}},\ }\href@noop {}
  {\bibfield  {journal} {\bibinfo  {journal} {Nature}\ }\textbf {\bibinfo
  {volume} {566}},\ \bibinfo {pages} {475} (\bibinfo {year}
  {2019})}\BibitemShut {NoStop}%
\bibitem [{\citenamefont {He}\ \emph {et~al.}(2019)\citenamefont {He},
  \citenamefont {Jiang}, \citenamefont {Zhang}, \citenamefont {Huang},
  \citenamefont {Fang},\ and\ \citenamefont {Jin}}]{he2019symtopo}%
  \BibitemOpen
  \bibfield  {author} {\bibinfo {author} {\bibfnamefont {Y.}~\bibnamefont
  {He}}, \bibinfo {author} {\bibfnamefont {Y.}~\bibnamefont {Jiang}}, \bibinfo
  {author} {\bibfnamefont {T.}~\bibnamefont {Zhang}}, \bibinfo {author}
  {\bibfnamefont {H.}~\bibnamefont {Huang}}, \bibinfo {author} {\bibfnamefont
  {C.}~\bibnamefont {Fang}}, \ and\ \bibinfo {author} {\bibfnamefont
  {Z.}~\bibnamefont {Jin}},\ }\href@noop {} {\bibfield  {journal} {\bibinfo
  {journal} {Chinese Physics B}\ }\textbf {\bibinfo {volume} {28}},\ \bibinfo
  {pages} {087102} (\bibinfo {year} {2019})}\BibitemShut {NoStop}%
\bibitem [{\citenamefont {Vergniory}\ \emph {et~al.}(2019)\citenamefont
  {Vergniory}, \citenamefont {Elcoro}, \citenamefont {Felser}, \citenamefont
  {Regnault}, \citenamefont {Bernevig},\ and\ \citenamefont
  {Wang}}]{vergniory2019complete}%
  \BibitemOpen
  \bibfield  {author} {\bibinfo {author} {\bibfnamefont {M.}~\bibnamefont
  {Vergniory}}, \bibinfo {author} {\bibfnamefont {L.}~\bibnamefont {Elcoro}},
  \bibinfo {author} {\bibfnamefont {C.}~\bibnamefont {Felser}}, \bibinfo
  {author} {\bibfnamefont {N.}~\bibnamefont {Regnault}}, \bibinfo {author}
  {\bibfnamefont {B.~A.}\ \bibnamefont {Bernevig}}, \ and\ \bibinfo {author}
  {\bibfnamefont {Z.}~\bibnamefont {Wang}},\ }\href@noop {} {\bibfield
  {journal} {\bibinfo  {journal} {Nature}\ }\textbf {\bibinfo {volume} {566}},\
  \bibinfo {pages} {480} (\bibinfo {year} {2019})}\BibitemShut {NoStop}%
\bibitem [{\citenamefont {Tang}\ \emph {et~al.}(2019)\citenamefont {Tang},
  \citenamefont {Po}, \citenamefont {Vishwanath},\ and\ \citenamefont
  {Wan}}]{tang2019comprehensive}%
  \BibitemOpen
  \bibfield  {author} {\bibinfo {author} {\bibfnamefont {F.}~\bibnamefont
  {Tang}}, \bibinfo {author} {\bibfnamefont {H.~C.}\ \bibnamefont {Po}},
  \bibinfo {author} {\bibfnamefont {A.}~\bibnamefont {Vishwanath}}, \ and\
  \bibinfo {author} {\bibfnamefont {X.}~\bibnamefont {Wan}},\ }\href@noop {}
  {\bibfield  {journal} {\bibinfo  {journal} {Nature}\ }\textbf {\bibinfo
  {volume} {566}},\ \bibinfo {pages} {486} (\bibinfo {year}
  {2019})}\BibitemShut {NoStop}%
\bibitem [{\citenamefont {Choudhary}\ \emph {et~al.}(2019)\citenamefont
  {Choudhary}, \citenamefont {Garrity},\ and\ \citenamefont
  {Tavazza}}]{choudhary2019high}%
  \BibitemOpen
  \bibfield  {author} {\bibinfo {author} {\bibfnamefont {K.}~\bibnamefont
  {Choudhary}}, \bibinfo {author} {\bibfnamefont {K.~F.}\ \bibnamefont
  {Garrity}}, \ and\ \bibinfo {author} {\bibfnamefont {F.}~\bibnamefont
  {Tavazza}},\ }\href@noop {} {\bibfield  {journal} {\bibinfo  {journal}
  {Scientific reports}\ }\textbf {\bibinfo {volume} {9}},\ \bibinfo {pages} {1}
  (\bibinfo {year} {2019})}\BibitemShut {NoStop}%
\bibitem [{\citenamefont {Gao}\ \emph {et~al.}(2020)\citenamefont {Gao},
  \citenamefont {Qian}, \citenamefont {Nie}, \citenamefont {Wang},
  \citenamefont {Weng},\ and\ \citenamefont {Fang}}]{gao2020high}%
  \BibitemOpen
  \bibfield  {author} {\bibinfo {author} {\bibfnamefont {J.}~\bibnamefont
  {Gao}}, \bibinfo {author} {\bibfnamefont {Y.}~\bibnamefont {Qian}}, \bibinfo
  {author} {\bibfnamefont {S.}~\bibnamefont {Nie}}, \bibinfo {author}
  {\bibfnamefont {Z.}~\bibnamefont {Wang}}, \bibinfo {author} {\bibfnamefont
  {H.}~\bibnamefont {Weng}}, \ and\ \bibinfo {author} {\bibfnamefont
  {Z.}~\bibnamefont {Fang}},\ }\href@noop {} {\bibfield  {journal} {\bibinfo
  {journal} {arXiv preprint arXiv:2004.09489}\ } (\bibinfo {year}
  {2020})}\BibitemShut {NoStop}%
\bibitem [{\citenamefont {Huang}\ \emph
  {et~al.}(2015{\natexlab{b}})\citenamefont {Huang}, \citenamefont {Xu},
  \citenamefont {Belopolski}, \citenamefont {Lee}, \citenamefont {Chang},
  \citenamefont {Wang}, \citenamefont {Alidoust}, \citenamefont {Bian},
  \citenamefont {Neupane}, \citenamefont {Zhang} \emph
  {et~al.}}]{huang2015weyl}%
  \BibitemOpen
  \bibfield  {author} {\bibinfo {author} {\bibfnamefont {S.-M.}\ \bibnamefont
  {Huang}}, \bibinfo {author} {\bibfnamefont {S.-Y.}\ \bibnamefont {Xu}},
  \bibinfo {author} {\bibfnamefont {I.}~\bibnamefont {Belopolski}}, \bibinfo
  {author} {\bibfnamefont {C.-C.}\ \bibnamefont {Lee}}, \bibinfo {author}
  {\bibfnamefont {G.}~\bibnamefont {Chang}}, \bibinfo {author} {\bibfnamefont
  {B.}~\bibnamefont {Wang}}, \bibinfo {author} {\bibfnamefont {N.}~\bibnamefont
  {Alidoust}}, \bibinfo {author} {\bibfnamefont {G.}~\bibnamefont {Bian}},
  \bibinfo {author} {\bibfnamefont {M.}~\bibnamefont {Neupane}}, \bibinfo
  {author} {\bibfnamefont {C.}~\bibnamefont {Zhang}},  \emph {et~al.},\
  }\href@noop {} {\bibfield  {journal} {\bibinfo  {journal} {Nature
  communications}\ }\textbf {\bibinfo {volume} {6}},\ \bibinfo {pages} {1}
  (\bibinfo {year} {2015}{\natexlab{b}})}\BibitemShut {NoStop}%
\bibitem [{\citenamefont {Lv}\ \emph {et~al.}(2015)\citenamefont {Lv},
  \citenamefont {Weng}, \citenamefont {Fu}, \citenamefont {Wang}, \citenamefont
  {Miao}, \citenamefont {Ma}, \citenamefont {Richard}, \citenamefont {Huang},
  \citenamefont {Zhao}, \citenamefont {Chen} \emph
  {et~al.}}]{lv2015experimental}%
  \BibitemOpen
  \bibfield  {author} {\bibinfo {author} {\bibfnamefont {B.}~\bibnamefont
  {Lv}}, \bibinfo {author} {\bibfnamefont {H.}~\bibnamefont {Weng}}, \bibinfo
  {author} {\bibfnamefont {B.}~\bibnamefont {Fu}}, \bibinfo {author}
  {\bibfnamefont {X.~P.}\ \bibnamefont {Wang}}, \bibinfo {author}
  {\bibfnamefont {H.}~\bibnamefont {Miao}}, \bibinfo {author} {\bibfnamefont
  {J.}~\bibnamefont {Ma}}, \bibinfo {author} {\bibfnamefont {P.}~\bibnamefont
  {Richard}}, \bibinfo {author} {\bibfnamefont {X.}~\bibnamefont {Huang}},
  \bibinfo {author} {\bibfnamefont {L.}~\bibnamefont {Zhao}}, \bibinfo {author}
  {\bibfnamefont {G.}~\bibnamefont {Chen}},  \emph {et~al.},\ }\href@noop {}
  {\bibfield  {journal} {\bibinfo  {journal} {Physical Review X}\ }\textbf
  {\bibinfo {volume} {5}},\ \bibinfo {pages} {031013} (\bibinfo {year}
  {2015})}\BibitemShut {NoStop}%
\bibitem [{\citenamefont {Weng}\ \emph {et~al.}(2015)\citenamefont {Weng},
  \citenamefont {Fang}, \citenamefont {Fang}, \citenamefont {Bernevig},\ and\
  \citenamefont {Dai}}]{weng2015weyl}%
  \BibitemOpen
  \bibfield  {author} {\bibinfo {author} {\bibfnamefont {H.}~\bibnamefont
  {Weng}}, \bibinfo {author} {\bibfnamefont {C.}~\bibnamefont {Fang}}, \bibinfo
  {author} {\bibfnamefont {Z.}~\bibnamefont {Fang}}, \bibinfo {author}
  {\bibfnamefont {B.~A.}\ \bibnamefont {Bernevig}}, \ and\ \bibinfo {author}
  {\bibfnamefont {X.}~\bibnamefont {Dai}},\ }\href@noop {} {\bibfield
  {journal} {\bibinfo  {journal} {Physical Review X}\ }\textbf {\bibinfo
  {volume} {5}},\ \bibinfo {pages} {011029} (\bibinfo {year}
  {2015})}\BibitemShut {NoStop}%
\bibitem [{\citenamefont {Prandini}\ \emph {et~al.}(2018)\citenamefont
  {Prandini}, \citenamefont {Marrazzo}, \citenamefont {Castelli}, \citenamefont
  {Mounet},\ and\ \citenamefont {Marzari}}]{prandini2018precision}%
  \BibitemOpen
  \bibfield  {author} {\bibinfo {author} {\bibfnamefont {G.}~\bibnamefont
  {Prandini}}, \bibinfo {author} {\bibfnamefont {A.}~\bibnamefont {Marrazzo}},
  \bibinfo {author} {\bibfnamefont {I.~E.}\ \bibnamefont {Castelli}}, \bibinfo
  {author} {\bibfnamefont {N.}~\bibnamefont {Mounet}}, \ and\ \bibinfo {author}
  {\bibfnamefont {N.}~\bibnamefont {Marzari}},\ }\href@noop {} {\bibfield
  {journal} {\bibinfo  {journal} {npj Computational Materials}\ }\textbf
  {\bibinfo {volume} {4}},\ \bibinfo {pages} {1} (\bibinfo {year}
  {2018})}\BibitemShut {NoStop}%
\bibitem [{\citenamefont {Hamann}(2013)}]{hamann2013optimized}%
  \BibitemOpen
  \bibfield  {author} {\bibinfo {author} {\bibfnamefont {D.}~\bibnamefont
  {Hamann}},\ }\href@noop {} {\bibfield  {journal} {\bibinfo  {journal}
  {Physical Review B}\ }\textbf {\bibinfo {volume} {88}},\ \bibinfo {pages}
  {085117} (\bibinfo {year} {2013})}\BibitemShut {NoStop}%
\bibitem [{\citenamefont {Bergerhoff}\ \emph {et~al.}(1987)\citenamefont
  {Bergerhoff}, \citenamefont {Brown}, \citenamefont {Allen} \emph
  {et~al.}}]{bergerhoff1987crystallographic}%
  \BibitemOpen
  \bibfield  {author} {\bibinfo {author} {\bibfnamefont {G.}~\bibnamefont
  {Bergerhoff}}, \bibinfo {author} {\bibfnamefont {I.}~\bibnamefont {Brown}},
  \bibinfo {author} {\bibfnamefont {F.}~\bibnamefont {Allen}},  \emph
  {et~al.},\ }\href@noop {} {\bibfield  {journal} {\bibinfo  {journal}
  {International Union of Crystallography, Chester}\ }\textbf {\bibinfo
  {volume} {360}},\ \bibinfo {pages} {77} (\bibinfo {year} {1987})}\BibitemShut
  {NoStop}%
\bibitem [{\citenamefont {Zagorac}\ \emph {et~al.}(2019)\citenamefont
  {Zagorac}, \citenamefont {M{\"u}ller}, \citenamefont {Ruehl}, \citenamefont
  {Zagorac},\ and\ \citenamefont {Rehme}}]{zagorac2019recent}%
  \BibitemOpen
  \bibfield  {author} {\bibinfo {author} {\bibfnamefont {D.}~\bibnamefont
  {Zagorac}}, \bibinfo {author} {\bibfnamefont {H.}~\bibnamefont {M{\"u}ller}},
  \bibinfo {author} {\bibfnamefont {S.}~\bibnamefont {Ruehl}}, \bibinfo
  {author} {\bibfnamefont {J.}~\bibnamefont {Zagorac}}, \ and\ \bibinfo
  {author} {\bibfnamefont {S.}~\bibnamefont {Rehme}},\ }\href@noop {}
  {\bibfield  {journal} {\bibinfo  {journal} {Journal of applied
  crystallography}\ }\textbf {\bibinfo {volume} {52}},\ \bibinfo {pages} {918}
  (\bibinfo {year} {2019})}\BibitemShut {NoStop}%
\bibitem [{\citenamefont {Gra{\v{z}}ulis}\ \emph {et~al.}(2012)\citenamefont
  {Gra{\v{z}}ulis}, \citenamefont {Da{\v{s}}kevi{\v{c}}}, \citenamefont
  {Merkys}, \citenamefont {Chateigner}, \citenamefont {Lutterotti},
  \citenamefont {Quiros}, \citenamefont {Serebryanaya}, \citenamefont {Moeck},
  \citenamefont {Downs},\ and\ \citenamefont {Le~Bail}}]{COD}%
  \BibitemOpen
  \bibfield  {author} {\bibinfo {author} {\bibfnamefont {S.}~\bibnamefont
  {Gra{\v{z}}ulis}}, \bibinfo {author} {\bibfnamefont {A.}~\bibnamefont
  {Da{\v{s}}kevi{\v{c}}}}, \bibinfo {author} {\bibfnamefont {A.}~\bibnamefont
  {Merkys}}, \bibinfo {author} {\bibfnamefont {D.}~\bibnamefont {Chateigner}},
  \bibinfo {author} {\bibfnamefont {L.}~\bibnamefont {Lutterotti}}, \bibinfo
  {author} {\bibfnamefont {M.}~\bibnamefont {Quiros}}, \bibinfo {author}
  {\bibfnamefont {N.~R.}\ \bibnamefont {Serebryanaya}}, \bibinfo {author}
  {\bibfnamefont {P.}~\bibnamefont {Moeck}}, \bibinfo {author} {\bibfnamefont
  {R.~T.}\ \bibnamefont {Downs}}, \ and\ \bibinfo {author} {\bibfnamefont
  {A.}~\bibnamefont {Le~Bail}},\ }\href@noop {} {\bibfield  {journal} {\bibinfo
   {journal} {Nucleic acids research}\ }\textbf {\bibinfo {volume} {40}},\
  \bibinfo {pages} {D420} (\bibinfo {year} {2012})}\BibitemShut {NoStop}%
\bibitem [{\citenamefont {Hinuma}\ \emph {et~al.}(2017)\citenamefont {Hinuma},
  \citenamefont {Pizzi}, \citenamefont {Kumagai}, \citenamefont {Oba},\ and\
  \citenamefont {Tanaka}}]{hinuma2017band}%
  \BibitemOpen
  \bibfield  {author} {\bibinfo {author} {\bibfnamefont {Y.}~\bibnamefont
  {Hinuma}}, \bibinfo {author} {\bibfnamefont {G.}~\bibnamefont {Pizzi}},
  \bibinfo {author} {\bibfnamefont {Y.}~\bibnamefont {Kumagai}}, \bibinfo
  {author} {\bibfnamefont {F.}~\bibnamefont {Oba}}, \ and\ \bibinfo {author}
  {\bibfnamefont {I.}~\bibnamefont {Tanaka}},\ }\href@noop {} {\bibfield
  {journal} {\bibinfo  {journal} {Computational Materials Science}\ }\textbf
  {\bibinfo {volume} {128}},\ \bibinfo {pages} {140} (\bibinfo {year}
  {2017})}\BibitemShut {NoStop}%
\bibitem [{\citenamefont {Gresch}\ \emph {et~al.}(2017)\citenamefont {Gresch},
  \citenamefont {Autes}, \citenamefont {Yazyev}, \citenamefont {Troyer},
  \citenamefont {Vanderbilt}, \citenamefont {Bernevig},\ and\ \citenamefont
  {Soluyanov}}]{gresch2017z2pack}%
  \BibitemOpen
  \bibfield  {author} {\bibinfo {author} {\bibfnamefont {D.}~\bibnamefont
  {Gresch}}, \bibinfo {author} {\bibfnamefont {G.}~\bibnamefont {Autes}},
  \bibinfo {author} {\bibfnamefont {O.~V.}\ \bibnamefont {Yazyev}}, \bibinfo
  {author} {\bibfnamefont {M.}~\bibnamefont {Troyer}}, \bibinfo {author}
  {\bibfnamefont {D.}~\bibnamefont {Vanderbilt}}, \bibinfo {author}
  {\bibfnamefont {B.~A.}\ \bibnamefont {Bernevig}}, \ and\ \bibinfo {author}
  {\bibfnamefont {A.~A.}\ \bibnamefont {Soluyanov}},\ }\href@noop {} {\bibfield
   {journal} {\bibinfo  {journal} {Physical Review B}\ }\textbf {\bibinfo
  {volume} {95}},\ \bibinfo {pages} {075146} (\bibinfo {year}
  {2017})}\BibitemShut {NoStop}%
\bibitem [{\citenamefont {Pizzi}\ \emph {et~al.}(2020)\citenamefont {Pizzi},
  \citenamefont {Vitale}, \citenamefont {Arita}, \citenamefont {Bl{\"u}gel},
  \citenamefont {Freimuth}, \citenamefont {G{\'e}ranton}, \citenamefont
  {Gibertini}, \citenamefont {Gresch}, \citenamefont {Johnson}, \citenamefont
  {Koretsune} \emph {et~al.}}]{pizzi2020wannier90}%
  \BibitemOpen
  \bibfield  {author} {\bibinfo {author} {\bibfnamefont {G.}~\bibnamefont
  {Pizzi}}, \bibinfo {author} {\bibfnamefont {V.}~\bibnamefont {Vitale}},
  \bibinfo {author} {\bibfnamefont {R.}~\bibnamefont {Arita}}, \bibinfo
  {author} {\bibfnamefont {S.}~\bibnamefont {Bl{\"u}gel}}, \bibinfo {author}
  {\bibfnamefont {F.}~\bibnamefont {Freimuth}}, \bibinfo {author}
  {\bibfnamefont {G.}~\bibnamefont {G{\'e}ranton}}, \bibinfo {author}
  {\bibfnamefont {M.}~\bibnamefont {Gibertini}}, \bibinfo {author}
  {\bibfnamefont {D.}~\bibnamefont {Gresch}}, \bibinfo {author} {\bibfnamefont
  {C.}~\bibnamefont {Johnson}}, \bibinfo {author} {\bibfnamefont
  {T.}~\bibnamefont {Koretsune}},  \emph {et~al.},\ }\href@noop {} {\bibfield
  {journal} {\bibinfo  {journal} {Journal of Physics: Condensed Matter}\
  }\textbf {\bibinfo {volume} {32}},\ \bibinfo {pages} {165902} (\bibinfo
  {year} {2020})}\BibitemShut {NoStop}%
\bibitem [{\citenamefont {Soluyanov}\ and\ \citenamefont
  {Vanderbilt}(2011)}]{soluyanov2011computing}%
  \BibitemOpen
  \bibfield  {author} {\bibinfo {author} {\bibfnamefont {A.~A.}\ \bibnamefont
  {Soluyanov}}\ and\ \bibinfo {author} {\bibfnamefont {D.}~\bibnamefont
  {Vanderbilt}},\ }\href@noop {} {\bibfield  {journal} {\bibinfo  {journal}
  {Physical Review B}\ }\textbf {\bibinfo {volume} {83}},\ \bibinfo {pages}
  {235401} (\bibinfo {year} {2011})}\BibitemShut {NoStop}%
\bibitem [{\citenamefont {Pizzi}\ \emph {et~al.}(2016)\citenamefont {Pizzi},
  \citenamefont {Cepellotti}, \citenamefont {Sabatini}, \citenamefont
  {Marzari},\ and\ \citenamefont {Kozinsky}}]{pizzi2016aiida}%
  \BibitemOpen
  \bibfield  {author} {\bibinfo {author} {\bibfnamefont {G.}~\bibnamefont
  {Pizzi}}, \bibinfo {author} {\bibfnamefont {A.}~\bibnamefont {Cepellotti}},
  \bibinfo {author} {\bibfnamefont {R.}~\bibnamefont {Sabatini}}, \bibinfo
  {author} {\bibfnamefont {N.}~\bibnamefont {Marzari}}, \ and\ \bibinfo
  {author} {\bibfnamefont {B.}~\bibnamefont {Kozinsky}},\ }\href@noop {}
  {\bibfield  {journal} {\bibinfo  {journal} {Computational Materials Science}\
  }\textbf {\bibinfo {volume} {111}},\ \bibinfo {pages} {218} (\bibinfo {year}
  {2016})}\BibitemShut {NoStop}%
\bibitem [{\citenamefont {Huber}\ \emph {et~al.}(2020)\citenamefont {Huber},
  \citenamefont {Zoupanos}, \citenamefont {Uhrin}, \citenamefont {Talirz},
  \citenamefont {Kahle}, \citenamefont {H{\"a}uselmann}, \citenamefont
  {Gresch}, \citenamefont {M{\"u}ller}, \citenamefont {Yakutovich},
  \citenamefont {Andersen} \emph {et~al.}}]{huber2020aiida}%
  \BibitemOpen
  \bibfield  {author} {\bibinfo {author} {\bibfnamefont {S.}~\bibnamefont
  {Huber}}, \bibinfo {author} {\bibfnamefont {S.}~\bibnamefont {Zoupanos}},
  \bibinfo {author} {\bibfnamefont {M.}~\bibnamefont {Uhrin}}, \bibinfo
  {author} {\bibfnamefont {L.}~\bibnamefont {Talirz}}, \bibinfo {author}
  {\bibfnamefont {L.}~\bibnamefont {Kahle}}, \bibinfo {author} {\bibfnamefont
  {R.}~\bibnamefont {H{\"a}uselmann}}, \bibinfo {author} {\bibfnamefont
  {D.}~\bibnamefont {Gresch}}, \bibinfo {author} {\bibfnamefont
  {T.}~\bibnamefont {M{\"u}ller}}, \bibinfo {author} {\bibfnamefont {A.~V.}\
  \bibnamefont {Yakutovich}}, \bibinfo {author} {\bibfnamefont {C.~W.}\
  \bibnamefont {Andersen}},  \emph {et~al.},\ }\href@noop {} {\bibfield
  {journal} {\bibinfo  {journal} {arXiv preprint arXiv:2003.12476}\ } (\bibinfo
  {year} {2020})}\BibitemShut {NoStop}%
\bibitem [{\citenamefont {Talirz}\ \emph {et~al.}(2020)\citenamefont {Talirz},
  \citenamefont {Kumbhar}, \citenamefont {Passaro}, \citenamefont {Yakutovich},
  \citenamefont {Granata}, \citenamefont {Gargiulo}, \citenamefont {Borelli},
  \citenamefont {Uhrin}, \citenamefont {Huber}, \citenamefont {Zoupanos} \emph
  {et~al.}}]{talirz2020materials}%
  \BibitemOpen
  \bibfield  {author} {\bibinfo {author} {\bibfnamefont {L.}~\bibnamefont
  {Talirz}}, \bibinfo {author} {\bibfnamefont {S.}~\bibnamefont {Kumbhar}},
  \bibinfo {author} {\bibfnamefont {E.}~\bibnamefont {Passaro}}, \bibinfo
  {author} {\bibfnamefont {A.~V.}\ \bibnamefont {Yakutovich}}, \bibinfo
  {author} {\bibfnamefont {V.}~\bibnamefont {Granata}}, \bibinfo {author}
  {\bibfnamefont {F.}~\bibnamefont {Gargiulo}}, \bibinfo {author}
  {\bibfnamefont {M.}~\bibnamefont {Borelli}}, \bibinfo {author} {\bibfnamefont
  {M.}~\bibnamefont {Uhrin}}, \bibinfo {author} {\bibfnamefont {S.~P.}\
  \bibnamefont {Huber}}, \bibinfo {author} {\bibfnamefont {S.}~\bibnamefont
  {Zoupanos}},  \emph {et~al.},\ }\href@noop {} {\bibfield  {journal} {\bibinfo
   {journal} {Scientific data}\ }\textbf {\bibinfo {volume} {7}},\ \bibinfo
  {pages} {299} (\bibinfo {year} {2020})}\BibitemShut {NoStop}%
\bibitem [{\citenamefont {Grassano}\ \emph {et~al.}(2023)\citenamefont
  {Grassano}, \citenamefont {Marzari},\ and\ \citenamefont
  {Campi}}]{mcloud2023weyl}%
  \BibitemOpen
  \bibfield  {author} {\bibinfo {author} {\bibfnamefont {D.}~\bibnamefont
  {Grassano}}, \bibinfo {author} {\bibfnamefont {N.}~\bibnamefont {Marzari}}, \
  and\ \bibinfo {author} {\bibfnamefont {D.}~\bibnamefont {Campi}},\
  }\href@noop {} {\enquote {\bibinfo {title} {{High-throughput screening of
  Weyl semimetals}},}\ }\bibinfo {howpublished}
  {\url{https://doi.org/10.24435/materialscloud:na-1b}} (\bibinfo {year}
  {2023})\BibitemShut {NoStop}%
\bibitem [{\citenamefont {Als{\'e}n}(1925)}]{alsen1925rontgenographische}%
  \BibitemOpen
  \bibfield  {author} {\bibinfo {author} {\bibfnamefont {N.}~\bibnamefont
  {Als{\'e}n}},\ }\href@noop {} {\bibfield  {journal} {\bibinfo  {journal}
  {Geologiska F{\"o}reningen i Stockholm F{\"o}rhandlingar}\ }\textbf {\bibinfo
  {volume} {47}},\ \bibinfo {pages} {19} (\bibinfo {year} {1925})}\BibitemShut
  {NoStop}%
\bibitem [{\citenamefont {Graf}\ \emph {et~al.}(2009)\citenamefont {Graf},
  \citenamefont {Casper}, \citenamefont {Winterlik}, \citenamefont {Balke},
  \citenamefont {Fecher},\ and\ \citenamefont {Felser}}]{graf2009crystal}%
  \BibitemOpen
  \bibfield  {author} {\bibinfo {author} {\bibfnamefont {T.}~\bibnamefont
  {Graf}}, \bibinfo {author} {\bibfnamefont {F.}~\bibnamefont {Casper}},
  \bibinfo {author} {\bibfnamefont {J.}~\bibnamefont {Winterlik}}, \bibinfo
  {author} {\bibfnamefont {B.}~\bibnamefont {Balke}}, \bibinfo {author}
  {\bibfnamefont {G.~H.}\ \bibnamefont {Fecher}}, \ and\ \bibinfo {author}
  {\bibfnamefont {C.}~\bibnamefont {Felser}},\ }\href@noop {} {\bibfield
  {journal} {\bibinfo  {journal} {Zeitschrift f{\"u}r anorganische und
  allgemeine Chemie}\ }\textbf {\bibinfo {volume} {635}},\ \bibinfo {pages}
  {976} (\bibinfo {year} {2009})}\BibitemShut {NoStop}%
\bibitem [{\citenamefont {Shi}\ \emph {et~al.}(2018)\citenamefont {Shi},
  \citenamefont {Muechler}, \citenamefont {Manna}, \citenamefont {Zhang},
  \citenamefont {Koepernik}, \citenamefont {Car}, \citenamefont {Van
  Den~Brink}, \citenamefont {Felser},\ and\ \citenamefont
  {Sun}}]{shi2018prediction}%
  \BibitemOpen
  \bibfield  {author} {\bibinfo {author} {\bibfnamefont {W.}~\bibnamefont
  {Shi}}, \bibinfo {author} {\bibfnamefont {L.}~\bibnamefont {Muechler}},
  \bibinfo {author} {\bibfnamefont {K.}~\bibnamefont {Manna}}, \bibinfo
  {author} {\bibfnamefont {Y.}~\bibnamefont {Zhang}}, \bibinfo {author}
  {\bibfnamefont {K.}~\bibnamefont {Koepernik}}, \bibinfo {author}
  {\bibfnamefont {R.}~\bibnamefont {Car}}, \bibinfo {author} {\bibfnamefont
  {J.}~\bibnamefont {Van Den~Brink}}, \bibinfo {author} {\bibfnamefont
  {C.}~\bibnamefont {Felser}}, \ and\ \bibinfo {author} {\bibfnamefont
  {Y.}~\bibnamefont {Sun}},\ }\href@noop {} {\bibfield  {journal} {\bibinfo
  {journal} {Physical Review B}\ }\textbf {\bibinfo {volume} {97}},\ \bibinfo
  {pages} {060406} (\bibinfo {year} {2018})}\BibitemShut {NoStop}%
\bibitem [{\citenamefont {Liu}\ \emph {et~al.}(2014)\citenamefont {Liu},
  \citenamefont {Li}, \citenamefont {Rajput}, \citenamefont {Gilks},
  \citenamefont {Lari}, \citenamefont {Galindo}, \citenamefont {Weinert},
  \citenamefont {Lazarov},\ and\ \citenamefont {Li}}]{liu2014tuning}%
  \BibitemOpen
  \bibfield  {author} {\bibinfo {author} {\bibfnamefont {Y.}~\bibnamefont
  {Liu}}, \bibinfo {author} {\bibfnamefont {Y.}~\bibnamefont {Li}}, \bibinfo
  {author} {\bibfnamefont {S.}~\bibnamefont {Rajput}}, \bibinfo {author}
  {\bibfnamefont {D.}~\bibnamefont {Gilks}}, \bibinfo {author} {\bibfnamefont
  {L.}~\bibnamefont {Lari}}, \bibinfo {author} {\bibfnamefont {P.}~\bibnamefont
  {Galindo}}, \bibinfo {author} {\bibfnamefont {M.}~\bibnamefont {Weinert}},
  \bibinfo {author} {\bibfnamefont {V.}~\bibnamefont {Lazarov}}, \ and\
  \bibinfo {author} {\bibfnamefont {L.}~\bibnamefont {Li}},\ }\href@noop {}
  {\bibfield  {journal} {\bibinfo  {journal} {Nature Physics}\ }\textbf
  {\bibinfo {volume} {10}},\ \bibinfo {pages} {294} (\bibinfo {year}
  {2014})}\BibitemShut {NoStop}%
\bibitem [{\citenamefont {Drummond}\ \emph {et~al.}(2012)\citenamefont
  {Drummond}, \citenamefont {Zolyomi},\ and\ \citenamefont
  {Fal'Ko}}]{Drummond.2012}%
  \BibitemOpen
  \bibfield  {author} {\bibinfo {author} {\bibfnamefont {N.~D.}\ \bibnamefont
  {Drummond}}, \bibinfo {author} {\bibfnamefont {V.}~\bibnamefont {Zolyomi}}, \
  and\ \bibinfo {author} {\bibfnamefont {V.~I.}\ \bibnamefont {Fal'Ko}},\
  }\href@noop {} {\bibfield  {journal} {\bibinfo  {journal} {Phys. Rev. B}\
  }\textbf {\bibinfo {volume} {85}},\ \bibinfo {pages} {075423} (\bibinfo
  {year} {2012})}\BibitemShut {NoStop}%
\bibitem [{\citenamefont {Grassano}\ \emph
  {et~al.}(2018{\natexlab{a}})\citenamefont {Grassano}, \citenamefont {Pulci},
  \citenamefont {Shubnyi}, \citenamefont {Sharapov}, \citenamefont {Gusynin},
  \citenamefont {Kavokin},\ and\ \citenamefont
  {Varlamov}}]{grassano2018detection}%
  \BibitemOpen
  \bibfield  {author} {\bibinfo {author} {\bibfnamefont {D.}~\bibnamefont
  {Grassano}}, \bibinfo {author} {\bibfnamefont {O.}~\bibnamefont {Pulci}},
  \bibinfo {author} {\bibfnamefont {V.}~\bibnamefont {Shubnyi}}, \bibinfo
  {author} {\bibfnamefont {S.}~\bibnamefont {Sharapov}}, \bibinfo {author}
  {\bibfnamefont {V.}~\bibnamefont {Gusynin}}, \bibinfo {author} {\bibfnamefont
  {A.}~\bibnamefont {Kavokin}}, \ and\ \bibinfo {author} {\bibfnamefont
  {A.}~\bibnamefont {Varlamov}},\ }\href@noop {} {\bibfield  {journal}
  {\bibinfo  {journal} {Physical Review B}\ }\textbf {\bibinfo {volume} {97}},\
  \bibinfo {pages} {205442} (\bibinfo {year} {2018}{\natexlab{a}})}\BibitemShut
  {NoStop}%
\bibitem [{\citenamefont {Menshchikova}\ \emph {et~al.}(2013)\citenamefont
  {Menshchikova}, \citenamefont {Otrokov}, \citenamefont {Tsirkin},
  \citenamefont {Samorokov}, \citenamefont {Bebneva}, \citenamefont {Ernst},
  \citenamefont {Kuznetsov},\ and\ \citenamefont
  {Chulkov}}]{menshchikova2013band}%
  \BibitemOpen
  \bibfield  {author} {\bibinfo {author} {\bibfnamefont {T.~V.}\ \bibnamefont
  {Menshchikova}}, \bibinfo {author} {\bibfnamefont {M.}~\bibnamefont
  {Otrokov}}, \bibinfo {author} {\bibfnamefont {S.}~\bibnamefont {Tsirkin}},
  \bibinfo {author} {\bibfnamefont {D.}~\bibnamefont {Samorokov}}, \bibinfo
  {author} {\bibfnamefont {V.}~\bibnamefont {Bebneva}}, \bibinfo {author}
  {\bibfnamefont {A.}~\bibnamefont {Ernst}}, \bibinfo {author} {\bibfnamefont
  {V.}~\bibnamefont {Kuznetsov}}, \ and\ \bibinfo {author} {\bibfnamefont
  {E.~V.}\ \bibnamefont {Chulkov}},\ }\href@noop {} {\bibfield  {journal}
  {\bibinfo  {journal} {Nano letters}\ }\textbf {\bibinfo {volume} {13}},\
  \bibinfo {pages} {6064} (\bibinfo {year} {2013})}\BibitemShut {NoStop}%
\bibitem [{\citenamefont {Edmonds}\ \emph {et~al.}(2016)\citenamefont
  {Edmonds}, \citenamefont {Hellerstedt}, \citenamefont {O’Donnell},
  \citenamefont {Tadich},\ and\ \citenamefont {Fuhrer}}]{edmonds2016molecular}%
  \BibitemOpen
  \bibfield  {author} {\bibinfo {author} {\bibfnamefont {M.~T.}\ \bibnamefont
  {Edmonds}}, \bibinfo {author} {\bibfnamefont {J.}~\bibnamefont
  {Hellerstedt}}, \bibinfo {author} {\bibfnamefont {K.~M.}\ \bibnamefont
  {O’Donnell}}, \bibinfo {author} {\bibfnamefont {A.}~\bibnamefont {Tadich}},
  \ and\ \bibinfo {author} {\bibfnamefont {M.~S.}\ \bibnamefont {Fuhrer}},\
  }\href@noop {} {\bibfield  {journal} {\bibinfo  {journal} {ACS applied
  materials \& interfaces}\ }\textbf {\bibinfo {volume} {8}},\ \bibinfo {pages}
  {16412} (\bibinfo {year} {2016})}\BibitemShut {NoStop}%
\bibitem [{\citenamefont {Grassano}\ \emph {et~al.}(2019)\citenamefont
  {Grassano}, \citenamefont {Pulci}, \citenamefont {Cannuccia},\ and\
  \citenamefont {Bechstedt}}]{grassano2019influence}%
  \BibitemOpen
  \bibfield  {author} {\bibinfo {author} {\bibfnamefont {D.}~\bibnamefont
  {Grassano}}, \bibinfo {author} {\bibfnamefont {O.}~\bibnamefont {Pulci}},
  \bibinfo {author} {\bibfnamefont {E.}~\bibnamefont {Cannuccia}}, \ and\
  \bibinfo {author} {\bibfnamefont {F.}~\bibnamefont {Bechstedt}},\ }\href@noop
  {} {\bibfield  {journal} {\bibinfo  {journal} {arXiv preprint
  arXiv:1906.12231}\ } (\bibinfo {year} {2019})}\BibitemShut {NoStop}%
\bibitem [{\citenamefont {Wang}\ \emph {et~al.}(2017)\citenamefont {Wang},
  \citenamefont {Sun}, \citenamefont {Lu},\ and\ \citenamefont
  {Xie}}]{wang_3d_2017}%
  \BibitemOpen
  \bibfield  {author} {\bibinfo {author} {\bibfnamefont {C.}~\bibnamefont
  {Wang}}, \bibinfo {author} {\bibfnamefont {H.-P.}\ \bibnamefont {Sun}},
  \bibinfo {author} {\bibfnamefont {H.-Z.}\ \bibnamefont {Lu}}, \ and\ \bibinfo
  {author} {\bibfnamefont {X.}~\bibnamefont {Xie}},\ }\href {\doibase
  10.1103/PhysRevLett.119.136806} {\bibfield  {journal} {\bibinfo  {journal}
  {Physical Review Letters}\ }\textbf {\bibinfo {volume} {119}},\ \bibinfo
  {pages} {136806} (\bibinfo {year} {2017})},\ \bibinfo {note} {publisher:
  American Physical Society}\BibitemShut {NoStop}%
\bibitem [{\citenamefont {Yang}\ \emph {et~al.}(2011)\citenamefont {Yang},
  \citenamefont {Lu},\ and\ \citenamefont {Ran}}]{yang_quantum_2011}%
  \BibitemOpen
  \bibfield  {author} {\bibinfo {author} {\bibfnamefont {K.-Y.}\ \bibnamefont
  {Yang}}, \bibinfo {author} {\bibfnamefont {Y.-M.}\ \bibnamefont {Lu}}, \ and\
  \bibinfo {author} {\bibfnamefont {Y.}~\bibnamefont {Ran}},\ }\href {\doibase
  10.1103/PhysRevB.84.075129} {\bibfield  {journal} {\bibinfo  {journal}
  {Physical Review B}\ }\textbf {\bibinfo {volume} {84}},\ \bibinfo {pages}
  {075129} (\bibinfo {year} {2011})},\ \bibinfo {note} {publisher: American
  Physical Society}\BibitemShut {NoStop}%
\bibitem [{\citenamefont {Grassano}\ \emph
  {et~al.}(2018{\natexlab{b}})\citenamefont {Grassano}, \citenamefont {Pulci},
  \citenamefont {Conte},\ and\ \citenamefont
  {Bechstedt}}]{grassano2018validity}%
  \BibitemOpen
  \bibfield  {author} {\bibinfo {author} {\bibfnamefont {D.}~\bibnamefont
  {Grassano}}, \bibinfo {author} {\bibfnamefont {O.}~\bibnamefont {Pulci}},
  \bibinfo {author} {\bibfnamefont {A.~M.}\ \bibnamefont {Conte}}, \ and\
  \bibinfo {author} {\bibfnamefont {F.}~\bibnamefont {Bechstedt}},\ }\href@noop
  {} {\bibfield  {journal} {\bibinfo  {journal} {Scientific reports}\ }\textbf
  {\bibinfo {volume} {8}},\ \bibinfo {pages} {1} (\bibinfo {year}
  {2018}{\natexlab{b}})}\BibitemShut {NoStop}%
\bibitem [{\citenamefont {Lee}\ \emph {et~al.}(2015)\citenamefont {Lee},
  \citenamefont {Xu}, \citenamefont {Huang}, \citenamefont {Sanchez},
  \citenamefont {Belopolski}, \citenamefont {Chang}, \citenamefont {Bian},
  \citenamefont {Alidoust}, \citenamefont {Zheng}, \citenamefont {Neupane}
  \emph {et~al.}}]{lee2015fermi}%
  \BibitemOpen
  \bibfield  {author} {\bibinfo {author} {\bibfnamefont {C.-C.}\ \bibnamefont
  {Lee}}, \bibinfo {author} {\bibfnamefont {S.-Y.}\ \bibnamefont {Xu}},
  \bibinfo {author} {\bibfnamefont {S.-M.}\ \bibnamefont {Huang}}, \bibinfo
  {author} {\bibfnamefont {D.~S.}\ \bibnamefont {Sanchez}}, \bibinfo {author}
  {\bibfnamefont {I.}~\bibnamefont {Belopolski}}, \bibinfo {author}
  {\bibfnamefont {G.}~\bibnamefont {Chang}}, \bibinfo {author} {\bibfnamefont
  {G.}~\bibnamefont {Bian}}, \bibinfo {author} {\bibfnamefont {N.}~\bibnamefont
  {Alidoust}}, \bibinfo {author} {\bibfnamefont {H.}~\bibnamefont {Zheng}},
  \bibinfo {author} {\bibfnamefont {M.}~\bibnamefont {Neupane}},  \emph
  {et~al.},\ }\href@noop {} {\bibfield  {journal} {\bibinfo  {journal}
  {Physical Review B}\ }\textbf {\bibinfo {volume} {92}},\ \bibinfo {pages}
  {235104} (\bibinfo {year} {2015})}\BibitemShut {NoStop}%
\bibitem [{\citenamefont {Grassano}\ \emph {et~al.}(2022)\citenamefont
  {Grassano}, \citenamefont {Binci},\ and\ \citenamefont
  {Marzari}}]{grassano2022prediction}%
  \BibitemOpen
  \bibfield  {author} {\bibinfo {author} {\bibfnamefont {D.}~\bibnamefont
  {Grassano}}, \bibinfo {author} {\bibfnamefont {L.}~\bibnamefont {Binci}}, \
  and\ \bibinfo {author} {\bibfnamefont {N.}~\bibnamefont {Marzari}},\
  }\href@noop {} {\bibfield  {journal} {\bibinfo  {journal} {arXiv preprint
  arXiv:2208.11412}\ } (\bibinfo {year} {2022})}\BibitemShut {NoStop}%
\end{thebibliography}%

\end{document}